# Combined stellar structure and atmosphere models for massive stars

## II. Spectral evolution on the main sequence

D. Schaerer[1], A. de Koter[2], W. Schmutz[3], and A. Maeder[1]

[1] Geneva Observatory, CH-1290 Sauverny, Switzerland; e-mail: schaerer@scsun.unige.ch, maeder@scsun.unige.ch
[2] NASA/GSCF USRA, Greenbelt, MD 20771, USA; e-mail: alex@homie.gsfc.nasa.gov
[3] Institut für Astronomie, ETH Zentrum, CH-8092 Zürich, Switzerland; e-mail: schmutz@astro.phys.ethz.ch

**Abstract.** In Schaerer et al. (1995, Paper I) we have presented the first "combined stellar structure and atmosphere models" (*CoStar*) for massive stars, which consistently treat the entire mass loosing star from the center out to the outer region of the stellar wind. The models use up-to-date input physics and state-of-the-art techniques to model both the stellar interior and the spherically expanding non–LTE atmosphere. The atmosphere models include line blanketing for all elements from hydrogen to zinc.

The present publication covers the spectral evolution corresponding to the main sequence interior evolution discussed in Paper I. The *CoStar* results presented in this paper comprise: *(a)* flux distributions, from the EUV to the far IR, and the ionizing fluxes in the hydrogen and helium continua, *(b)* absolute optical and infrared UBVRIJHKLMN photometric magnitudes and UV colors, *(c)* detailed line blanketed UV spectra, and *(d)* non–LTE hydrogen and helium line spectra in the optical and IR, including theoretical K band spectra. These results may, e.g., be used for population synthesis models intended to study the massive star content in young starforming regions.

We compare our results with other predictions from LTE and non–LTE plane parallel models and point out the improvements and the importance of using adequate atmosphere models including stellar winds for massive stars. Particular emphasis is given to comparisons of the UV spectral evolution with observations, including continuum indices and several metal line signatures of P-Cygni lines and broad absorption features. Good agreement is found for most UV features. In particular, we are able to reproduce the strong observed Fe III 1920 Å feature in late O and early B giants and supergiants. This feature is found to depend sensitively on temperature and may be used to derive effective temperatures for these stars.

We also derive a simple formula to determine mass loss rates from the equivalent width of hydrogen recombination lines (H$\alpha$, P$\alpha$ and B$\alpha$) for OB stars showing net emission in one or more of these lines.

## 1. Introduction

This is the second paper on "combined stellar structure and atmosphere models" (*CoStar*). In *CoStar* models we treat the stellar interior, the line-blanketed non-LTE spherically extended atmosphere and the wind in its entirety. The models are intended to study massive stars ($M \gtrsim 20 M_\odot$) as for these stars mass loss through a stellar wind is important for the stellar evolution (cf. Chiosi & Maeder 1986, Maeder & Conti 1994) and also because of the impact of the winds of these stars on the emergent flux distribution (e.g. Kudritzki & Hummer 1991). *CoStar* models may therefore be used to study the continuum and (metal)line spectra of stars in evolutionary phases corresponding to OB, LBV, Of/WN, and Wolf–Rayet (WR) stars.

An important advantage and motive for this work is that our approach allows for a more 'direct' comparison of stellar evolution results with observations. A second incentive for this work is that our method allows us to study the effect of a spherically expanding atmosphere and stellar wind on the envelope, interior structure and evolution of massive stars. Last, but not least, our models e.g. WR winds and LBV outbursts, which are likely to require a consistent modeling of both the stellar envelope and the wind.

Our approach implies the consistent modeling of both the UV, optical and IR spectral ranges. A future aim is to derive stellar parameters from those diagnostics that are most sensitive to a specific parameter. This includes the determination of metal abundances from UV spectral line features. These UV features are at the same time likely to yield valuable constraints on the metal ionization, which is important for the dynamical modeling of stellar winds.

In the present work we concentrate on the main sequence (MS) and present the first study of the spectral evolution of the continuum and metal line spectrum (see Schaerer 1995ab for WR phases). Up to now, theoretical studies have either been limited to individual objects, or have concentrated on morphological studies of the most important P-Cygni lines (Pauldrach et al. 1990). The reason is directly related to the complex physical conditions prevailing in massive early-type stars. This in contrast to the situation for intermediate and low mass stars, where hydrostatic LTE models provide an good description (but see Najarro et al. 1995). To give an example of the status of diagnostics in early type stars: at present quantitative abundance determinations for OB stars are essentially limited to H and He (e.g. Herrero et al. 1992), while the first determinations including metals has only been achieved recently (Pauldrach et al. 1994).

The characteristics of our code are described in detail in Schaerer et al. (1995, hereafter paper I). The code takes the non–LTE conditions in a spherically expanding atmosphere into account. The photosphere and the wind are treated consistently in a unified way, i.e. we do not use the core-halo approximation (de Koter et al. 1993). In addition we include line blanketing following the opacity sampling method introduced by Schmutz (1991). A description of the sampling technique and a detailed analysis of the effects of line blanketing for OB stars is given by Schaerer & Schmutz (1994ab, hereafter SS94ab) and Schaerer (1995a).

Our theoretical approach encompasses many aspects of the spectral evolution of massive stars. Our model predictions include the flux distribution from the extreme-UV ($\sim 30$ Å) to the far-IR ($\sim 200$ μm) and yield detailed line profiles of the strongest H and He lines. We provide ionizing fluxes, which may be of interest for studies of young stellar populations, e.g. in clusters, H II regions, and starbursts. Synthetic UBVRIJHKLMN photometric data is derived. We present detailed predictions of the UV spectrum taking into account the spectral lines of all elements from H to Zn and we compare these spectra with IUE observations.

In Sect. 2 we first present the selected models describing the detailed spectral evolution. The continuum spectral energy distributions, including ionizing fluxes, are discussed, and we compare our results with observations. The evolution of the H and He line spectrum, from the optical to the IR, is discussed in Sect. 5, where we also present an analysis of hydrogen recombination lines which can be used as mass loss indicators (Sect. 5.1). The main conclusions are summarized in Sect. 6.

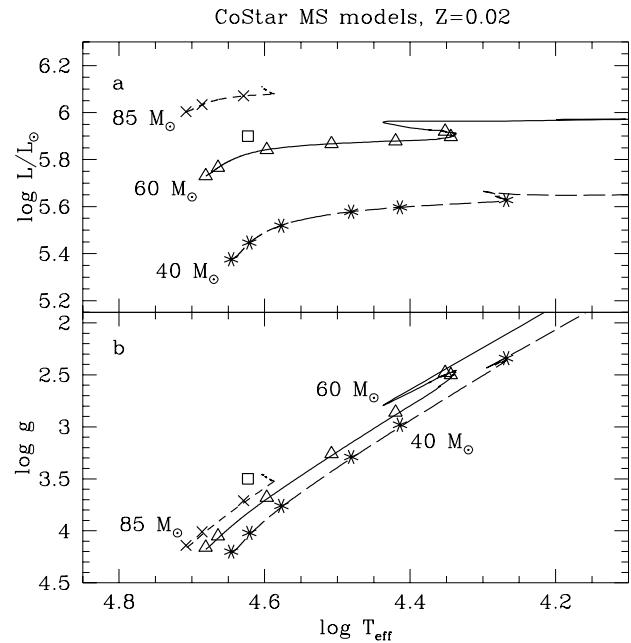

**Fig. 1. a** HR–diagram covering the MS phases for initial masses of 40, 60, and 85 $M_\odot$. The WR stage during the H–burning phase of the 85 $M_\odot$ model is not included. Crosses, triangles and stars denote the selected models describing the tracks; the square indicates the position of ζ Puppis (see Table 1). **b** $\log g$–$\log T_{\rm eff}$ diagram corresponding to the upper panel

## 2. Selected models

In this section we describe the spectral evolution along the evolutionary tracks for initial masses $M_i = 40, 60$, and 85 $M_\odot$. We have selected several models on each evolutionary path, for which we will discuss detailed results. The stellar parameters of these models are given in Table 1. Their position in the HR–diagram and the $\log g$–$\log T_{\rm eff}$ diagram is shown in Fig. 1. In this figure we have included the O4I(n)f star ζ Puppis, analysed in detail by Schaerer & Schmutz (1994ab).

The following entries are given in Table 1: Model number (column 1), age (2), present mass (3), the effective temperature (4), the luminosity (5), gravity (6), stellar radius (7), mass loss rate (8), terminal velocity (9), number fraction of hydrogen $n_{\rm H}$ (10), and helium $n_{\rm He}$ (11), and the mean molecular weight per free particle $\mu$ used to

**Table 1.** Summary of selected models: stellar parameters

| model # | age [yr] | $\frac{M}{M_\odot}$ | $\log T_{\rm eff}$ [K] | $\log \frac{L}{L_\odot}$ | $\log g$ [cm s$^{-2}$] | $\frac{R_\star}{R_\odot}$ | $\log \dot M$ [$M_\odot$yr$^{-1}$] | $v_\infty$ [km s$^{-1}$] | $n_{\rm H}$ | $n_{\rm He}$ | $\mu$ | SpType |
|---|---|---|---|---|---|---|---|---|---|---|---|---|
| 40 $M_\odot$ track: | | | | | | | | | | | | |
| 1 | 4.04 10$^4$ | 39.98 | 4.646 | 5.376 | 4.20 | 8.326 | -6.463 | 2962. | 0.90 | 0.10 | 0.634 | O5V |
| 2 | 1.50 10$^6$ | 39.25 | 4.621 | 5.447 | 4.02 | 10.152 | -6.170 | 2690. | 0.90 | 0.10 | 0.634 | O5.5IV |
| 3 | 2.85 10$^6$ | 37.86 | 4.577 | 5.519 | 3.76 | 13.457 | -5.809 | 2354. | 0.90 | 0.10 | 0.634 | O6.5II |
| 4 | 3.81 10$^6$ | 35.48 | 4.481 | 5.578 | 3.29 | 22.422 | -5.414 | 1893. | 0.90 | 0.10 | 0.634 | O9.5III |
| 5 | 4.08 10$^6$ | 34.35 | 4.414 | 5.597 | 2.98 | 31.215 | -5.290 | 1660. | 0.90 | 0.10 | 0.634 | B0I |
| 6 | 4.38 10$^6$ | 32.38 | 4.268 | 5.628 | 2.34 | 63.521 | -5.126 | 1269. | 0.90 | 0.10 | 0.634 | B2I |
| 60 $M_\odot$ track: | | | | | | | | | | | | |
| 7 | 8.72 10$^4$ | 59.96 | 4.681 | 5.731 | 4.16 | 10.663 | -6.090 | 3050. | 0.90 | 0.10 | 0.634 | O4V |
| 8 | 7.72 10$^5$ | 59.20 | 4.664 | 5.766 | 4.05 | 12.010 | -5.856 | 2883. | 0.90 | 0.10 | 0.634 | O4.5V |
| 9 | 2.23 10$^6$ | 55.03 | 4.597 | 5.842 | 3.68 | 17.838 | -5.252 | 2381. | 0.90 | 0.10 | 0.634 | O6.5V |
| 10 | 2.76 10$^6$ | 50.60 | 4.508 | 5.867 | 3.26 | 27.672 | -4.931 | 1960. | 0.90 | 0.10 | 0.634 | O9II |
| 11 | 3.00 10$^6$ | 47.43 | 4.420 | 5.879 | 2.86 | 42.147 | -4.826 | 1646. | 0.90 | 0.10 | 0.634 | B0I |
| 12 | 3.23 10$^6$ | 43.68 | 4.344 | 5.897 | 2.50 | 60.890 | -4.779 | 1391. | 0.87 | 0.13 | 0.666 | B1.5III |
| 13 | 3.44 10$^6$ | 40.01 | 4.352 | 5.920 | 2.48 | 60.392 | -4.747 | 1319. | 0.80 | 0.20 | 0.737 | B1II |
| 85 $M_\odot$ track: | | | | | | | | | | | | |
| 14 | 5.00 10$^4$ | 84.88 | 4.708 | 6.004 | 4.14 | 12.900 | -5.683 | 3182. | 0.90 | 0.10 | 0.634 | O3IV |
| 15 | 7.10 10$^5$ | 82.84 | 4.686 | 6.034 | 4.01 | 14.780 | -5.379 | 2977. | 0.90 | 0.10 | 0.634 | O3II |
| 16 | 1.66 10$^6$ | 75.58 | 4.629 | 6.071 | 3.71 | 20.017 | -4.879 | 2538. | 0.90 | 0.10 | 0.634 | O5III |
| $\zeta$ Puppis[1]: | | | | | | | | | | | | |
| | | 35. | 4.623 | 5.9 | 3.5 | 17. | -5.523 | 2200. | 0.80 | 0.20 | | O4 I(n)f |

[1]: Schaerer & Schmutz (1994a), model A

determine the photospheric structure (12). The last column gives an approximate spectral classification, which has been obtained from a nearest neighbour search in the tables of Schmidt-Kaler (1982) using the variables ($T_{\rm eff}, M_{\rm bol}$) giving a large weight to the temperature. If instead we determine the classification by nearest neighbour in the HR-diagram, i.e. ($\log T_{\rm eff}, \log L/L_\odot$), the resulting differences may reach up to two spectral subtypes and two luminosity classes. This should roughly indicate the precision of the assigned spectral classification.

## 3. Continuous spectrum

The evolution of the line blanketed flux distribution along the MS is shown in Fig. 2 for models along the 60 $M_\odot$ track. The bolometric luminosity increase from the ZAMS (model 7) to the TAMS (model 13) is a factor of 1.55. Note the progressive appearance of line blanketing features in the far UV. This will be discussed in detail in Sect. 4.

Figure 3 shows a comparison of the EUV to IR spectrum predicted by our non–LTE line blanketed wind models (solid line) with a non–LTE H&He plane parallel atmosphere (Herrero 1994, dotted) and a line blanketed LTE model from Kurucz (1991, dashed). The stellar parameters are chosen as close as possible to that of $\zeta$ Puppis (see Table 1): $T_{\rm eff}$=43 kK, $\log g$=3.7 for the non–LTE plane parallel atmosphere, and $T_{\rm eff}$=42 kK, $\log g$=4.5 for the Kurucz model. As expected, the energy distribution of all models is virtually identical from the near UV to the near IR. At wavelengths $\lambda \gtrsim 10$ $\mu$m, the spherically expanding atmosphere shows the characteristic flux enhancement due to free-free processes in the wind. The largest differences are found shortward of the He II edge ($\lambda < 228$ Å) where, compared to LTE and non–LTE plane parallel models, the flux is significantly enhanced due to the presence of the wind outflow (cf. Gabler et al. 1989). Similarly, non–LTE and partly also wind effects increase the flux shortward of the He I edge ($\lambda < 504$ Å). Non-LTE and wind effects even affect the Lyman continuum flux for lower temperatures. These important consequences for the ionizing flux will be discussed below.

### 3.1. Ionizing spectrum

In Table 2 we list the predicted number of photons emerging at wavelengths shorter than 912, 504, and 228 Å respectively, referred to as $q_0$, $q_1$, and $q_2$. The total ionizing luminosity $Q_i$ (in photons s$^{-1}$) is obtained by $Q_i = 4\pi(R_\star R_\odot)^2 q_i$, where $R_\star$ is given in Table 1.

It is difficult to estimate the reliability of the predicted ionizing fluxes, since they not only depend on the detailed atmospheric modeling (temperature structure, blanketing

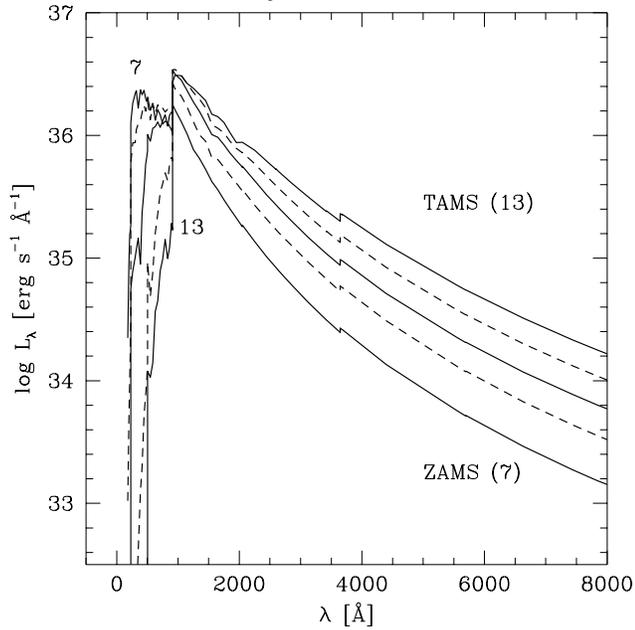
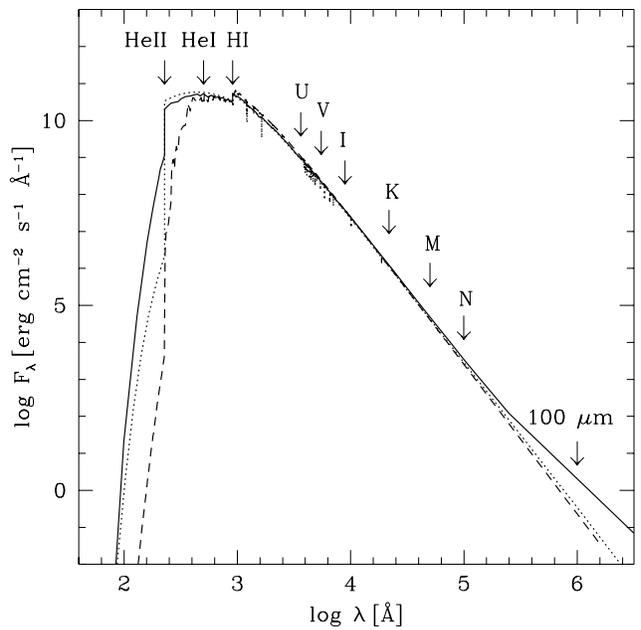

**Fig. 2.** Predicted line blanketed flux distributions during the MS evolution of a 60 $M_\odot$ star. Presented are models 7, 9, 10, 11, and 13 (cf. table 1). The spectra from model 8 and 12 are very similar to 7 and 13 respectively, and have therefore been excluded from this plot. Note that the model close to the ZAMS (7) has the largest flux at $\lambda < 912$ Å, and the lowest flux longward of the H Lyman edge. With decreasing temperature the strength of the blanketing features progressively increases in the observable UV spectral range from $\sim$ 1500 to 2000 Å

**Fig. 3.** Comparison of emergent spectra from our non–LTE line blanketed wind models (solid line) with a non–LTE plane parallel model (Herrero 1994, dotted), and a plane parallel LTE model of Kurucz (1991, ATLAS 9, dashed line). Also indicated are the H Lyman edge, the He I, and the He II Lyman edge, as well as the position of some photometric bands (Johnson). The stellar parameters are $T_{\rm eff}$=42 kK (43 kK for the dashed) and $\log g$ values of 3.5 to 4.5. The additional parameters for the wind model ($\dot M$, $v_\infty$) correspond to $\zeta$ Puppis (see Table 1)

etc.), but also on the wind parameters (see also Najarro et al. 1995). The values for $Q_2$ are affected by the largest uncertainty and they strongly depend on the wind density. In cases where the radiation shortwards of the He II edge originates from deep layers (near to the photosphere), the flux at $\lambda < 228$ Å is quite sensitive to the atmospheric density and temperature structure (cf. also Gabler et al. 1992). For example, according to whether the photospheric structure is calculated taking into account radiation pressure using OPAL opacities or using only electron scattering (cf. Paper I), the predicted He II ionizing flux is found to vary by up to 0.6 dex for the models close to the ZAMS. This number may be taken as an indication of the uncertainty of the $Q_2$ predictions. Another potential source of uncertainty for high $T_{\rm eff}$ models could come from the coherent treatment of electron scattering, which might modify the He ionization, as recently pointed out by Rybicki & Hummer (1994).

Figure 4 shows the temporal evolution of the ionizing luminosity along the tracks. During a large fraction of the H-burning lifetime, the ionizing luminosities $Q_0$ and $Q_1$ remain approximately constant and close to the ZAMS value. A constant $Q_0$ and $Q_1$ is in fact often assumed in calculations of integrated properties of stellar populations (e.g. Vacca 1994). The largest difference is obtained for the He$^0$ ionizing photons along the 40 $M_\odot$ track, where the mean value of $Q_1$ over the MS lifetime is approximately half the value at the ZAMS. The situation is significantly different for the He$^+$ ionizing photons. As soon as the stars have evolved to temperatures $T_{\rm eff} \lesssim 35$ kK, the wind becomes optically thick to radiation at $\lambda < 228$ Å, resulting in a rapid drop of the number of He$^+$ ionizing photons. This corresponds to ages of $\sim$ 2.5 to 3 Myr. In a young stellar population the presence of nebular lines with ionization potentials above the He$^+$ edge thus yields a strong upper limit on its age, provided that other ionization sources such as hot Wolf-Rayet stars or non-stellar sources (shock excitation, photoionization by X-rays) can be excluded (cf. Schmutz et al. 1992).

### 3.1.1. He$^+$ ionizing photons

In Fig. 5 we have plotted the ratio of He II to H ionizing photons as a function of effective temperature. The results are shown for our line blanketed models. To allow for a comparison, we also plot the values obtained without line blanketing. The spread of the differences between blanketed and non-blanketed models is due to differences

**Table 2.** Ionizing photon fluxes in cm$^{-2}$s$^{-1}$ from line blanketed models

| model | log $q_0$ | log $q_1$ | log $q_2$ |
|---|---|---|---|
| 40 $M_\odot$ track: | | | |
| 1 | 24.46 | 23.97 | 21.56 |
| 2 | 24.33 | 23.75 | 21.27 |
| 3 | 24.05 | 23.31 | 20.13 |
| 4 | 23.00 | 20.84 | 8.69 |
| 5 | 22.38 | 19.95 | 6.27 |
| 6 | 20.90 | 14.02 | |
| 60 $M_\odot$ track: | | | |
| 7 | 24.67 | 24.20 | 21.95 |
| 8 | 24.58 | 24.09 | 21.54 |
| 9 | 24.23 | 23.59 | 20.44 |
| 10 | 23.64 | 22.48 | 12.42 |
| 11 | 22.68 | 20.18 | 7.10 |
| 12 | 21.65 | 15.59 | |
| 13 | 21.74 | 15.62 | |
| 85 $M_\odot$ track: | | | |
| 14 | 24.81 | 24.39 | 21.55 |
| 15 | 24.71 | 24.25 | 21.50 |
| 16 | 24.45 | 23.88 | 20.89 |
| $\zeta$ Puppis: | | | |
| | 24.41 | 23.86 | 20.98 |

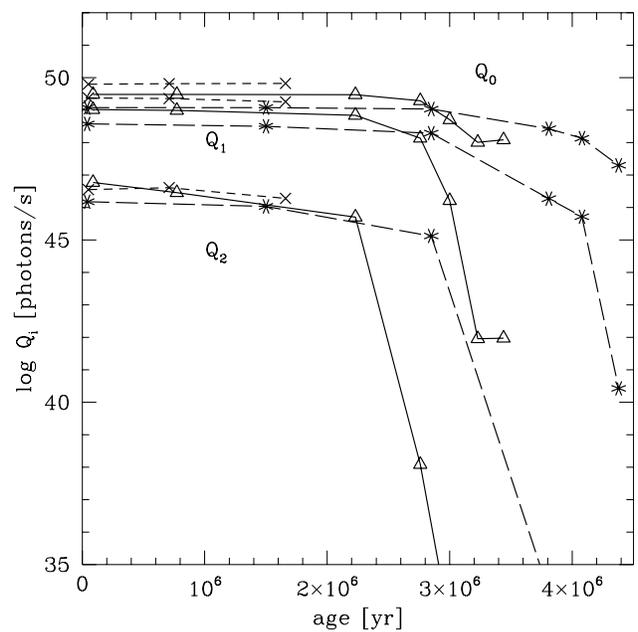

**Fig. 4.** Temporal evolution of the ionizing photon fluxes shortward of 912, 504, and 228 Å respectively. The short-dashed line connects the evolution of the 85 $M_\odot$ model (crosses), the solid line those of the 60 $M_\odot$ (triangles), and the long-dashed line the 40 $M_\odot$ model (stars). There are three groups of lines for $Q_0$, $Q_1$, and $Q_2$ respectively

in the calculation of the deep atmospheric structure (see above). As demonstrated by Schaerer & Schmutz (1994a) line blanketing leads to a higher He ionization and an increased radiation field at $\lambda < 228$ Å, which explains the larger $Q_2/Q_0$ ratios obtained for the blanketed models. Also given in the figure are the predictions from plane parallel non–LTE models for different gravities (Clegg & Middlemass 1987). The spherically extended and expanding atmosphere models predict approximately two to three orders of magnitudes more flux in the He II continuum than plane parallel atmospheres. As Gabler et al. (1989) have shown, this is due to the presence of the wind outflow. Our predictions for $Q_2$ are of the same order as the values from the unified models of Gabler et al. (1992). For precise comparisons, however, one has to remember that for OB stars the He II Lyman flux depends sensitively on the atmospheric structure (see above).

### 3.1.2. H ionizing photons

The number of Lyman continuum photons are compared with results from previous determinations in Fig. 6. Shown is the ionizing photon flux per unit surface area, which allows a direct comparison with predictions from plane parallel models independently of a radius calibration. We compare our models with the widely used results from Panagia (1973), the recent calculations from Voels et al. (1989) and those from Bohannan et al. (1990). The latter two results are based on plane parallel non–LTE models, which account for backscattered radiation from the wind. As expected, for temperatures above $T_{\rm eff} \sim 40$ kK, where the Lyman flux can be well approximated by a black-body spectrum, all predictions agree. We add that a good agreement is also found with the plane parallel wind blanketed models of Kudritzki et al. (1991), calculated for $35 \leq T_{\rm eff} \leq 51$ kK. For the most evolved models, the Lyman continuum flux is, approximately, up to a factor of two larger than the Panagia (1973) values — the exact value is in particular also dependent on the adopted value for $\dot{M}$. The enhancement of $q_0$ is due to the spherical extension and non–LTE effects related to the presence of the wind. The same result was found recently by Najarro et al. (1995), who discuss the physical effect in detail. They also point out the possible underestimate of the contribution of B stars to the diffuse ionizing radiation.

### 3.1.3. He$^\circ$ ionizing photons

Regarding the He$^\circ$ ionizing fluxes good agreement is obtained with the results of Kudritzki et al. (1991, plane parallel, wind blanketed non–LTE models): For the range covered by their models our predictions lie well between their luminosity classes V and I, and differ by less than $\sim 15$ %. As mentioned by Vacca (1994), the $q_1$ values obtained from non–LTE plane parallel atmospheres are sys-tematically larger (by factors of 1.6 to 3.7 between $T_{\rm eff}=35$ and 51 kK) than the values derived from Kurucz models (see e.g. Fig. 3). For $q_1$, similar words of caution regarding the mass loss dependence as for $q_0$ are appropriate.

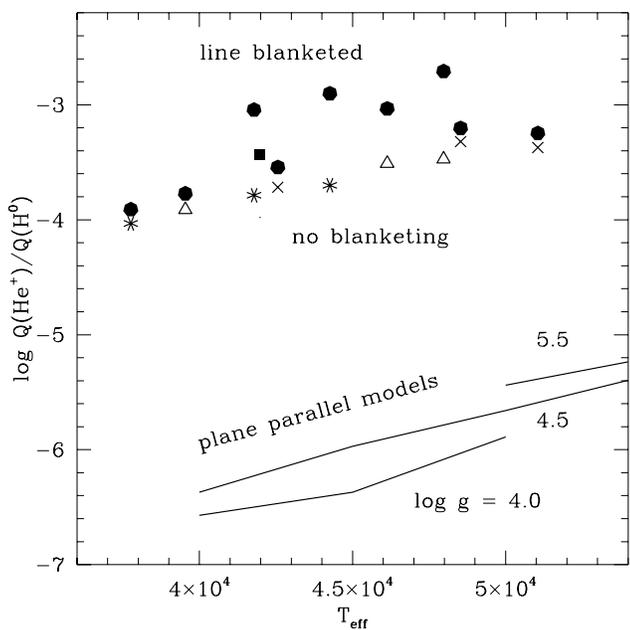
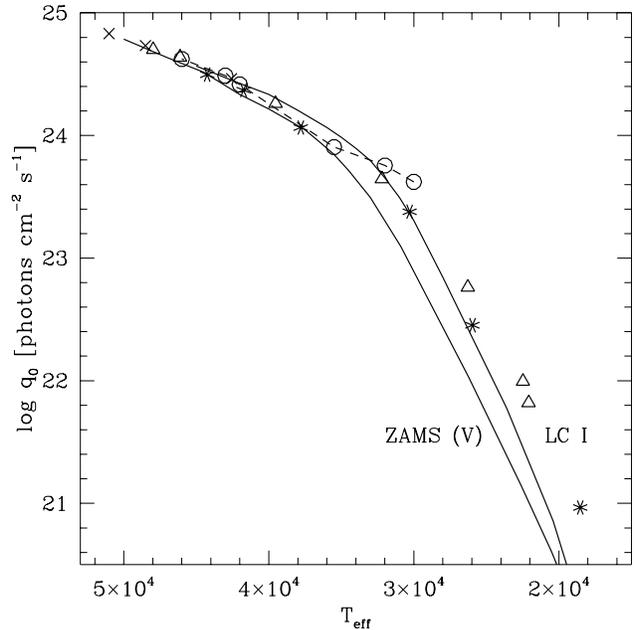

**Fig. 5.** Ratio of the He$^+$ to H ionizing photons, $Q_2/Q_1$, as a function of the effective temperature. Circles denote line blanketed models. The effect of line blanketing is illustrated by including the non blanketed models at the same $T_{\rm eff}$ (same symbols as in Fig. 4). The square denotes the blanketed $\zeta$ Puppis model from Schaerer & Schmutz (1994a). Solid lines give the values obtained from plane parallel non–LTE models of Clegg & Middlemass (1987) labeled with their $\log g$ value. The large increase of the He$^+$ ionizing flux is due to the presence of the wind

**Fig. 6.** Logarithm of the Lyman continuum flux (photons cm$^{-2}$s$^{-1}$) versus effective temperature. The predictions from our line blanketed atmosphere models are indicated as crosses for the 85 $M_\odot$ track, triangles (60 $M_\odot$ track), and stars (40 $M_\odot$ track). The solid lines shows the relations from Panagia (1973) for his ZAMS and luminosity class I. The dashed line connects the wind blanketed models from Voels et al. (1989) and Bohannan et al. (1990) (circles).

*3.2. Photometric evolution*

We have listed the absolute magnitudes for the Johnson UBVRIJKLMN system and the ESO H filter in Table 3. Following de Koter et al. (1995), we have calculated monochromatic absolute magnitudes using the calibration of Johnson (1966, $F_\nu$ values from Table IV), Lamla (1982, Table 39 for B), and Koornneef (1983) for the ESO filter. Note that the calibration of the B filter is quite uncertain, which can result in differences up to 0.$^{\rm m}$1 relative to other calibrations. To check the assumption of monochromatic magnitudes, we convolved our spectra with the response functions of Lamla (1982, Table 11, 37 for UBVRIJ). For ZAMS models the largest differences were found for RIJ ($\sim$ 0.$^{\rm m}$1–0.$^{\rm m}$15 smaller). Differences in U can amount to 0.$^{\rm m}$1, while B and V show smaller differences. Therefore, for more precise determinations the appropriate response functions should be used.

In particular our predictions of infra red (IR) magnitudes, taking the effect of the spherically expanding non–LTE atmosphere into account, may provide a useful base for many applications in studies of hot star populations. However, one should keep in mind that the IR flux depends on the adopted wind parameters, in particular on the wind density. As an example: for the supergiant model 12 an increase of $\dot M/v_\infty$ by a factor of 3.5 (8) yields fluxes, which differ by less than 0.$^{\rm m}$05 ($\sim$ 0.$^{\rm m}$1) for UBVRI, by 0.$^{\rm m}$1 (0.$^{\rm m}$5) for K, and by up to 0.$^{\rm m}$6 (1.$^{\rm m}$8) for the N filter. Our predictions should therefore be used with caution for stars with significantly different mass loss rates and/or wind velocities. For LBV's the influence of parameter changes on the emergent spectrum and on photometric quantities has been studied by de Koter et al. (1995).

*3.3. UV continuum indices*

The optical continuum of hot stars carries very little information about gravity and temperature because it is in the Rayleigh-Jeans domain of the spectrum. For photometric studies one is therefore led to use bands located in the UV spectral range.

In Table 4 we list the color indices from the continuum flux of our line blanketed models for UV passbands

**Table 3.** Monochromatic absolute magnitudes for the Johnson UBVRIJKLMN system. Also given is the ESO H magnitude

| model | U | B | V | R | I | J | K | L | M | N | H (ESO) |
|---|---|---|---|---|---|---|---|---|---|---|---|
| 40 $M_\odot$ track: | | | | | | | | | | | |
| 1 | -6.08 | -4.94 | -4.64 | -4.44 | -4.19 | -3.89 | -3.88 | -3.77 | -3.62 | -3.73 | -3.87 |
| 2 | -6.46 | -5.34 | -5.04 | -4.85 | -4.61 | -4.30 | -4.26 | -4.20 | -4.15 | -4.41 | -4.27 |
| 3 | -6.99 | -5.89 | -5.61 | -5.43 | -5.19 | -4.89 | -4.85 | -4.71 | -4.51 | -4.60 | -4.85 |
| 4 | -7.64 | -6.60 | -6.34 | -6.16 | -5.96 | -5.66 | -5.66 | -5.55 | -5.35 | -5.47 | -5.86 |
| 5 | -7.89 | -6.91 | -6.67 | -6.52 | -6.34 | -6.06 | -6.08 | -5.99 | -5.80 | -5.96 | -6.06 |
| 6 | -8.53 | -7.70 | -7.53 | -7.42 | -7.31 | -7.07 | -7.14 | -7.06 | -6.86 | -7.01 | -7.11 |
| 60 $M_\odot$ track: | | | | | | | | | | | |
| 7 | -6.77 | -5.61 | -5.31 | -5.11 | -4.86 | -4.55 | -4.51 | -4.37 | -4.14 | -4.20 | -4.51 |
| 8 | -6.97 | -5.83 | -5.54 | -5.34 | -5.11 | -4.83 | -4.80 | -4.67 | -4.49 | -4.66 | -4.81 |
| 9 | -7.59 | -6.47 | -6.19 | -6.00 | -5.77 | -5.47 | -5.46 | -5.34 | -5.15 | -5.31 | - 5.44 |
| 10 | -8.12 | -7.05 | -6.78 | -6.61 | -6.40 | -6.12 | -6.13 | -6.04 | -5.88 | -6.13 | -6.11 |
| 11 | -8.57 | -7.57 | -7.33 | -7.18 | -7.00 | -6.73 | -6.75 | -6.66 | -6.49 | -6.71 | -6.73 |
| 12 | -8.93 | -7.99 | -7.79 | -7.66 | -7.51 | -7.24 | -7.30 | -7.22 | -7.04 | -7.23 | -7.27 |
| 13 | -8.98 | -8.02 | -7.81 | -7.68 | -7.52 | -7.27 | -7.31 | -7.23 | -7.05 | -7.24 | -7.29 |
| 85 $M_\odot$ track: | | | | | | | | | | | |
| 14 | -7.25 | -6.09 | -5.79 | -5.59 | -5.34 | -5.02 | -4.99 | -4.86 | -4.66 | -4.75 | -4.99 |
| 15 | -7.46 | -6.31 | -6.02 | -5.82 | -5.57 | -5.26 | -5.23 | -5.12 | -4.94 | -5.02 | -5.23 |
| 16 | -7.87 | -6.75 | -6.46 | -6.27 | -6.04 | -5.75 | -5.80 | -5.72 | -5.54 | -5.80 | -5.75 |

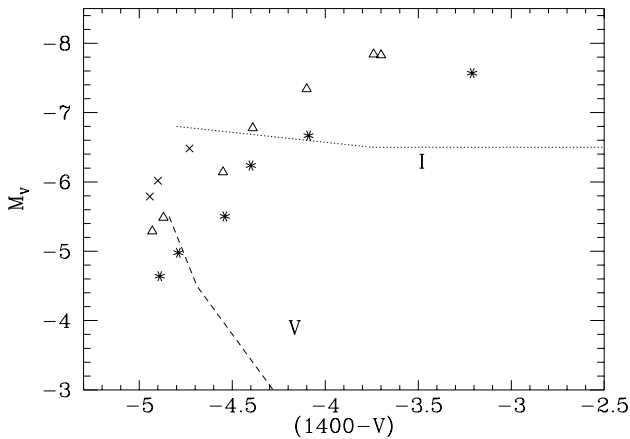

**Fig. 7.** UV color–magnitude diagram for the $(1400 - V)$ index. The absolute visual magnitude is taken from Table 3. Using UV colors, different evolutionary phases are clearly separated (symbols as in Fig. 1). The dashed and the dotted line show the relations for the mean values for dwarfs and supergiants respectively given by Fanelli et al. (1992). The large stretch in $(1400 - V)$ illustrates the power of UV photometry as a diagnostic tool for hot stars

as defined by Fanelli et al. (1992, hereafter FCBW). The UV bands are rectangular boxes of width $\Delta\lambda = 200$ Å, approximately centered at the given wavelength (in Å). Using IUE low resolution spectra FCBW have built mean spectra for a large number of groups of spectral types and luminosity classes (LC), which were then used to derive UV colors and line indices. We will compare our calculations with their results.

In Fig. 7 we plot our MS tracks in the $M_V$ vs $(1400-V)$ diagram. The $(1400 - V)$ color index varies by up to two orders of magnitude between spectral types $\sim$ O5 V and B2 I. This clearly illustrates the usefulness of UV indices as a diagnostic tool for hot stars. The mean observed values from FCBW for dwarfs and supergiants, up to the spectral type O4, are also shown in Fig. 7. Our models cover the entire observed range between LC V and I, the most extreme models being more luminous than the mean LC I. In Fig. 7 our ZAMS models appear to be $\sim 0\overset{m}{.}1$–$0\overset{m}{.}2$ too blue in $(1400 - V)$. However, the standard deviation (using equal weights) from the stars in the O3-6 V group of FCBW (see Fig. 7) is $0\overset{m}{.}23$. Thus, our results for the observed continuum slope between 1400 and 5500 Å agrees within one sigma. We add that our predicted $(1400 - V)$ colors also agree with those derived from Kurucz (1991) models.

To make a more detailed comparison of the predicted UV continua with observations we have plotted the $(1400 - 1700)$ vs. $(1700 - 2600)$ color–color in Fig. 8a. Again, the mean color indices from FCBW for dwarfs and supergiants of spectral types between O4 and B9 are superposed in this figure. As the majority of our models are roughly of spectral types O3 to O9, they are crowded on the left of the figure, where $(1400 - 1700) \sim -0\overset{m}{.}5$ and $(1700-2600) \sim -1\overset{m}{.}4$ to $-1\overset{m}{.}5$. This is in good agreement with the observations. Note also that for these models the predicted blanketing effects lead only to small changes of the colors considered above. Surprisingly however, the remaining models, corresponding to later types and supergiants, are located outside the region spanned by the observations. The systematic shifts indicates mainly that the observed $(1400-1700)$ color must be bluer than the models predict. A comparison with the $(1400-2600)$ vs. $(2600-V)$ diagram (Fig. 8b) shows a good agreement with the observations. This implies that the difference noted above must essentially be due to the 1700 band. Indeed, for the most discrepant supergiant model (model 6) a decrease of the flux in the 1700 band by only $0\overset{m}{.}15$ suffices to reconcile our models with the observed $(1400-1700)$ and $(1700-2600)$ colors for supergiants.

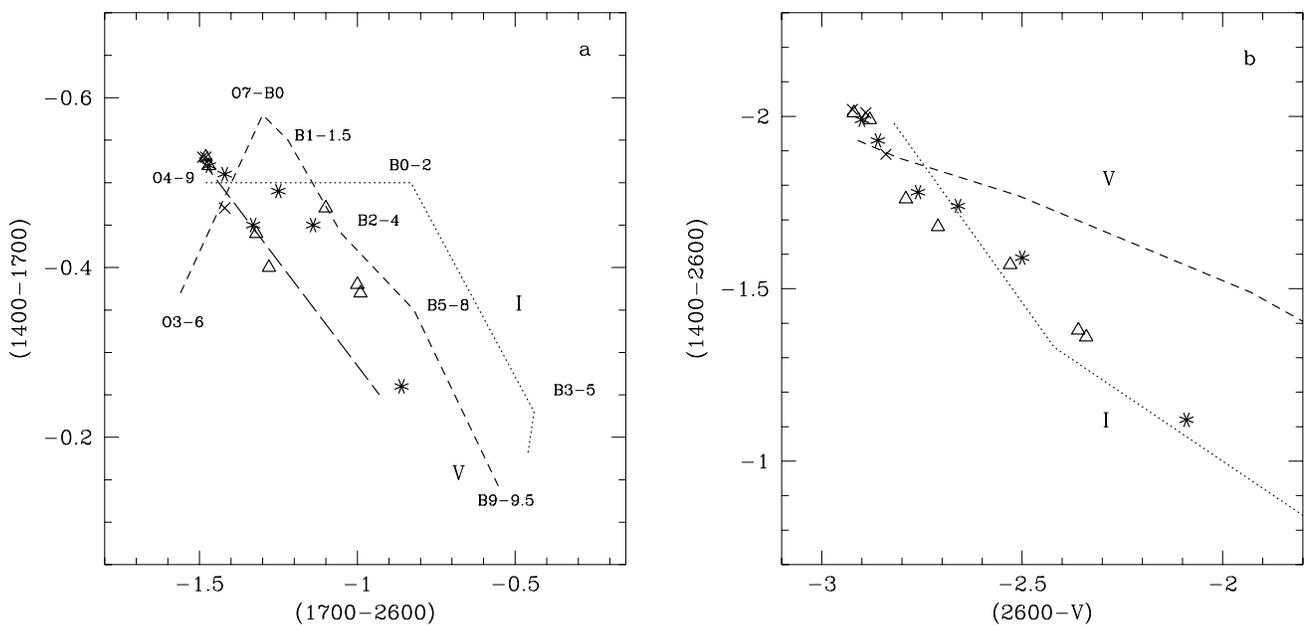

**Fig. 8. a** UV color–color diagram for the bands from FCBW defined in the text. Our models are shown by the same symbols as in Fig. 1. Note that the models of approximate luminosity class V, and IV (cf. Table 1) are indistinguishable and located at $(1400-1700) \sim -0\overset{m}{.}5$ and $(1700-2600) \sim -1\overset{m}{.}4$ to $-1\overset{m}{.}5$. The giant and supergiant models are well separated. The long-dashed line indicates the position of our models if we exclude line blanketing. Observations (mean colors from Fanelli et al. 1992) are shown for Dwarfs (dashed) and Supergiants (dotted). The labels along these mean relations indicate the position of the spectral type groups, which have been used to derive the relations. The most evolved models are found to have an excess of up to $\sim 0\overset{m}{.}15$ mag in the 1700 band.
**b** Same as left panel showing $(1400-2600)$ vs. $(2600-V)$. The models are in good agreement with the mean observed relation.

To investigate the cause of this difference, we compared our models with several IUE spectra of supergiants, as used by FCBW. This allows us to explain the discrepancy: The observed spectrum plotted in Fig. 9 (HD 165024, B2 Ib) is from the B0-2 I group. Contrary to our model, the observed spectrum has a steeper gradient between the 1400 and 1700 bands. As explained above this difference is essentially due to a larger flux depression in the 1700 band. The most likely explanation is that over a large fraction of the 1700 band numerous Fe IV lines (cf. Bruhweiler et al. 1981) provide important blocking, which is underestimated in our models. In fact, in this wavelength domain the supergiant spectra look very similar to the spectra of Of/WN objects, for which Schaerer (1995a) demonstrated that the large flux depression is mainly due to Fe IV. A more quantitative discussion of several line blanketing features is given below. We thus conclude that in the 1700 band our models yield an overestimate of up to 15 % in the flux, while for the other far–UV colors we find a general agreement with the observations.

## 4. Evolution of the UV line blanketed spectrum

Our line blanketed atmosphere models allow us to predict a large number of observable line features, which are very well suited for comparisons with IUE low resolution spectra. Before doing so, we again (see Paper I) like to stress that three important assumptions enter in the calculation of spectra by Monte-Carlo simulation. First and second, those on the ionization and excitation structure, as discussed in SS94ab. Third, those on the line forming mechanism. In the present work we concentrate on wind features since photospheric broadening is not treated. In our Monte-Carlo method, we include line scattering in the Sobolev approximation taking the continuum opacity into account. To be precise, collisional deexcitation and the effect of branching are not taken into account. The weakest point in our calculations is likely the neglect of line broadening yielding only a poor treatment of photospheric lines. We would like to stress that the problem of non-LTE line blanketing in hot star atmospheres is a very complicated one and a full treatment is as yet not feasible (see Pauldrach et al. 1994). Our calculations are a first step towards this goal.

**Table 4.** UV continuum color indices (in magnitudes)

| model | $1400 - V$ | $1700 - V$ | $2600 - V$ | $3000 - V$ |
|---|---|---|---|---|
| 40 $M_\odot$ track: | | | | |
| 1 | -4.89 | -4.37 | -2.90 | -2.27 |
| 2 | -4.79 | -4.28 | -2.86 | -2.24 |
| 3 | -4.54 | -4.09 | -2.76 | -2.17 |
| 4 | -4.40 | -3.91 | -2.66 | -1.99 |
| 5 | -4.09 | -3.64 | -2.50 | -1.97 |
| 6 | -3.21 | -2.95 | -2.09 | -1.65 |
| 60 $M_\odot$ track: | | | | |
| 7 | -4.93 | -4.40 | -2.92 | -2.29 |
| 8 | -4.87 | -4.35 | -2.88 | -2.26 |
| 9 | -4.55 | -4.11 | -2.79 | -2.20 |
| 10 | -4.39 | -3.99 | -2.71 | -2.12 |
| 11 | -4.10 | -3.63 | -2.53 | -1.99 |
| 12 | -3.70 | -3.33 | -2.34 | -1.85 |
| 13 | -3.74 | -3.36 | -2.36 | -1.87 |
| 85 $M_\odot$ track: | | | | |
| 14 | -4.94 | -4.41 | -2.92 | -2.29 |
| 15 | -4.90 | -4.37 | -2.89 | -2.28 |
| 16 | -4.73 | -4.26 | -2.84 | -2.23 |

Due to the adopted frequency binning, the spectral resolution varies nonuniformly. It is largest longward of the H Lyman edge at 912 Å and the He II Paschen edge at 2046 Å. The resolution ranges between $\lambda/\Delta\lambda \sim 100 - 1000$ with a typical value of 500. At all wavelengths the relative statistical error of the predicted flux is less than $10^{-3}$.

In the following, we will study the evolution of the most important wind lines, but we will also discuss other metal line features. The predicted UV spectral evolution for the $M_i = 40$, 60, and 85 $M_\odot$ tracks is shown in Figs. 10, 11 and 12 respectively. The illustrated spectral range reaches from $\sim 900$ to 2200 Å. The main sequence evolution approximately starts with spectral types O5 to O3 dwarfs, which then evolve to B2 to O5 supergiants (cf. Table 1).

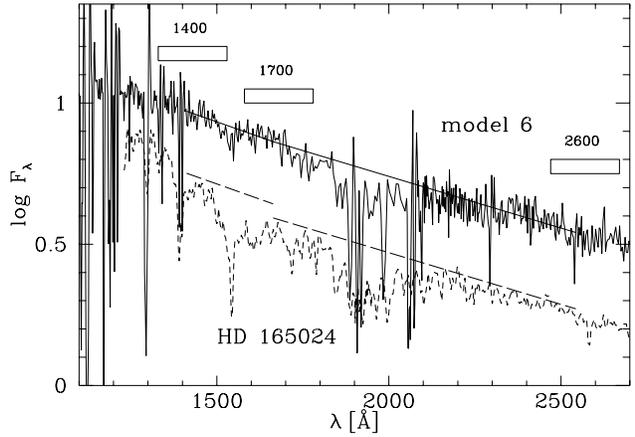

**Fig. 9.** Comparison of the predicted UV spectrum of model 6 (solid line) with IUE low resolution observations of HD 165024 (B2 Ib, dashed). The observed spectrum has been dereddened using $E(B-V)=0\overset{m}{.}1$ (FCBW) and the Seaton (1979) law. The scaling of both observations and the model spectrum are arbitrary. Also shown are the positions of the broad far-UV bands 1400, 1700, and 2600. The thin solid line gives the slope of the model spectrum between these bands (drawn by hand). The thin dashed line is obtained by shifting the $(1400 - 1700)$ and $(1700 - 2600)$ slopes of the model to the observed spectrum. The observations clearly show a bluer $(1400-1700)$ index probably due to important blending in the 1700 band

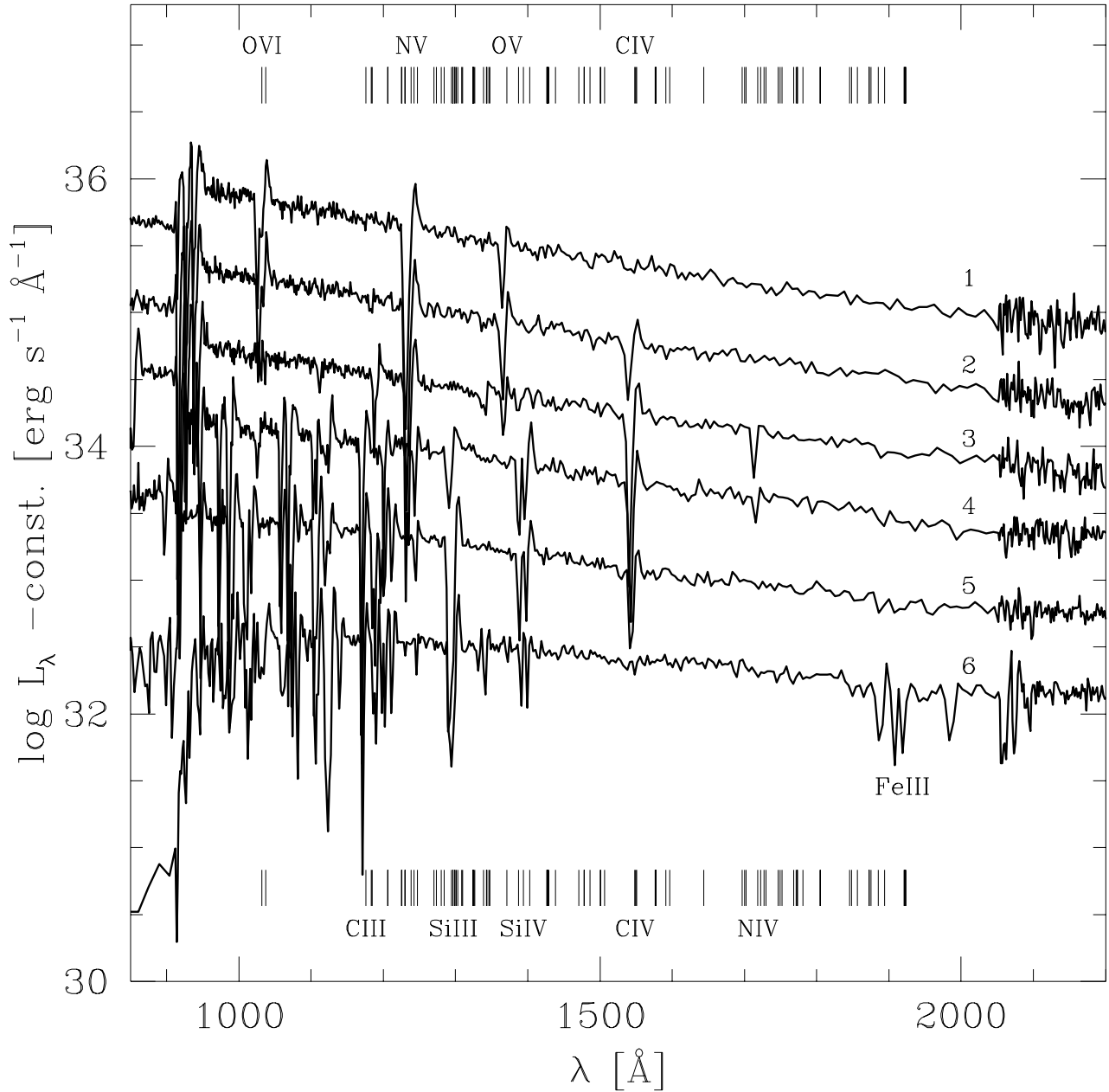

**Fig. 10.** Synthetic UV spectra showing the spectral evolution on the 40 $M_\odot$ track (models 1–6). Plotted is the logarithm of the emergent luminosity. The evolution along the MS proceeds from top (dwarf) to bottom (supergiant). Starting with the second model, each spectrum has been shifted downwards by 0.7 dex with respect to the previous one, in order to allow a good comparison. The marks on the top and bottom indicate the location of the CNO and Si lines taken from the lists of Bruhweiler et al. (1981) and Dean & Bruhweiler (1985). The strongest CNO, and Si features are labeled. Note also the strong Fe III 1920 Å feature in supergiants

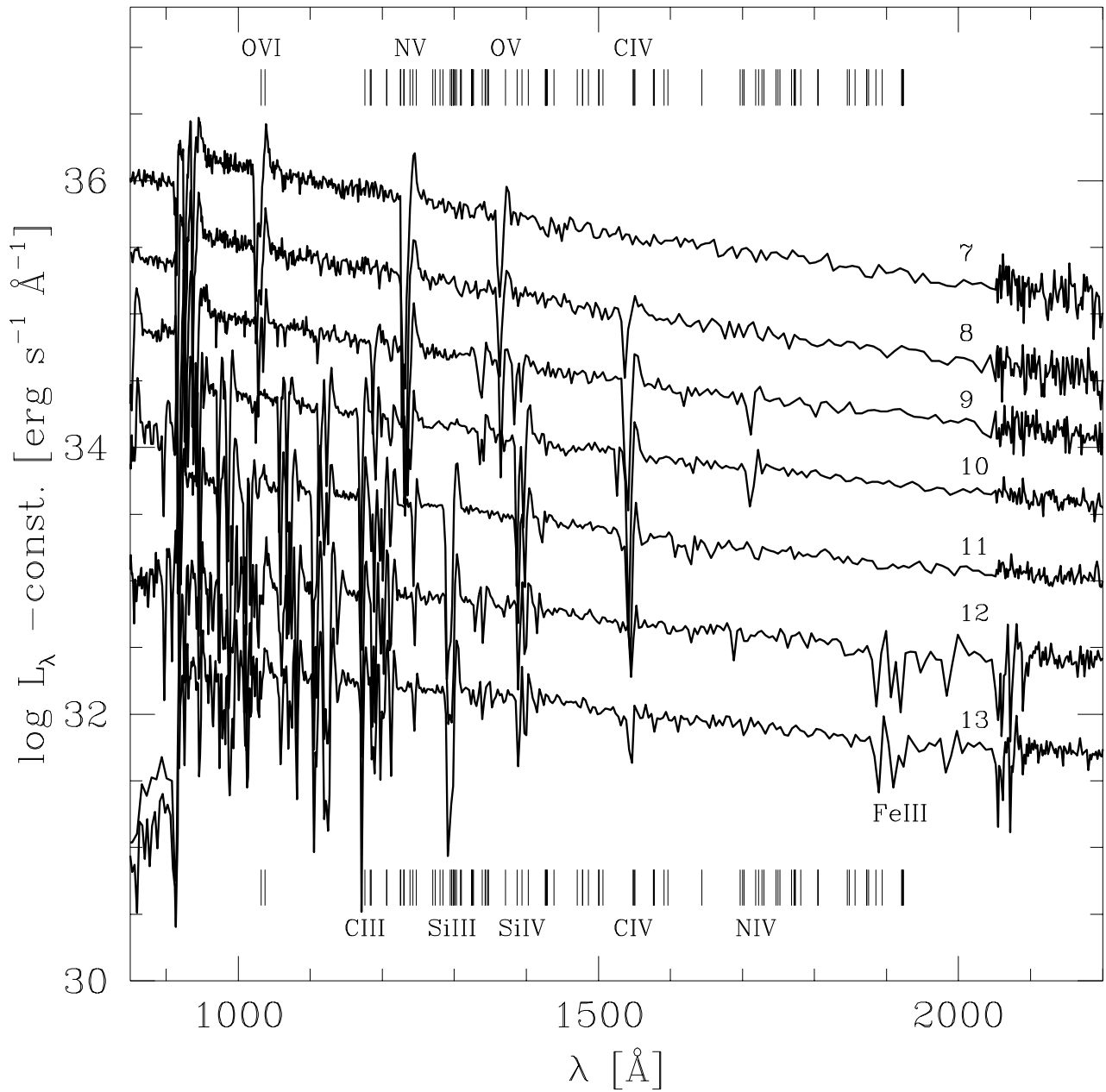

**Fig. 11.** Same as Fig. 10 for the 60 $M_\odot$ track (models 7 to 13)

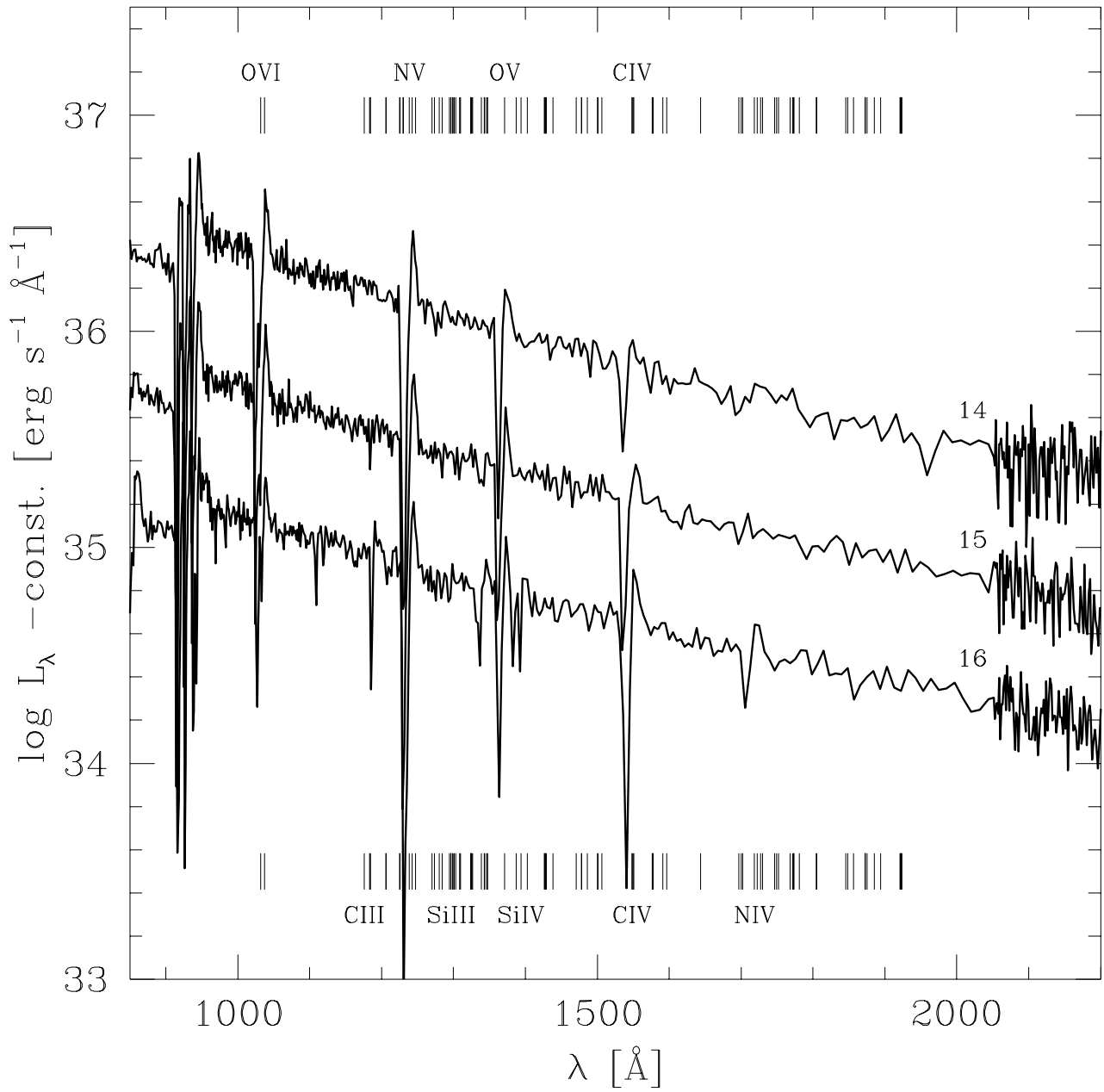

**Fig. 12.** Same as Fig. 10 for the 85 $M_\odot$ track (models 14 to 16)

O VI λλ 1032, 1038, N V λλ 1239, 1243, Si IV λλ 1394, 1403 and C IV λλ 1548, 1551 are clearly present in the synthetic spectra plotted in Figs. 10 to 12. A first inspection of the strong wind lines yields the following:

*(a)* The predictions for the N V resonance line follows the observed decrease in line strength towards later spectral types. The spectra of the B type stars, show some tendency to produce less N V than observed, which reflects the problem of super-ionization. The observed N IV λ 1718 line shows no clear trend in the domain of our calculations (Snow et al. 1994). The observed profiles are of different types including photospheric lines, shifted or asymmetric absorption and P-Cygni profiles. We find the same variety of profiles in our models.

*(b)* The strong luminosity dependence of the Si IV line is reproduced.

*(c)* The observed C IV line in O3-5 dwarfs is very strong. Our ZAMS models, however, do not produce this feature. We will return to this problem later on. For the domain covered by our models, the observations of the C III λ 1175 line show a gradual transitions from photospheric to P-Cygni profile (Snow et al. 1994). The appearance of this C III feature in our models follows the observed trend.

*(d)* We predict a rather strong O V λ 1371 P-Cygni line for the hottest models, while it is mostly observed in absorption. This might be due to an overestimate of either the ionization fraction of O V, or the excitation of the corresponding level. We note that for the hottest models, O V is indeed predicted to be the dominant ionization stage in contrast to the calculations of MacFarlane et al. (1993) and Pauldrach et al. (1994). Similarly the overionization of O is also apparent if we compare the O IV λ 1338 and O V λ 1371 lines with the observations compiled by Snow et al. (1994).

Let us now make a systematic and quantitative comparison of the predicted spectra with observations. We again use the results of FCBW, who introduced several far–UV spectral line indices which have been derived for a large number of stars using IUE low resolution spectra covering luminosity classes V to I and spectral types O3 to K or M.

In Fig. 13 we illustrate the position of the narrow bands of FCBW (bands 2–20), and of three bands defined by Nandy et al. (1981, called 54, 56 and 57). These bandpasses were chosen to include the most prominent UV absorption and P-Cygni features and adjacent "continuum" regions. The characteristics of the bands 54, 56, and 57 are given in Table 5. Spectral indices, derived from combinations of these bandpasses, quantify the strength of spectral lines. Line indices are determined by mean fluxes over three bandpasses, one being centered on the wavelength of the feature, and two on bracketing sidebands. All indices are measured in magnitudes. The indices of FCBW are calculated according to $I_i = -2.5 \log(F_\lambda^i / F_\lambda^c)$,

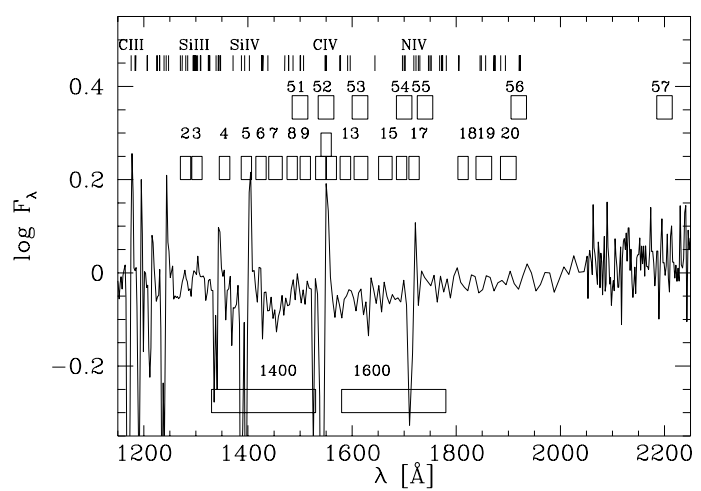

**Fig. 13.** Detail of normalised UV spectrum (model 10) to illustrate the position of the far UV narrow bands listed in Table 5. Bands 2–20 are from FCBW; 54, 56 and 57 from Nandy et al. . The position of two of the broad bands of FCBW is shown at the bottom. As in Fig. 10 the marks indicate the positions of observed CNO and Si lines

where $F_\lambda^i$ is the mean flux in the feature band $i$, and $F_\lambda^c$ is the linearly interpolated continuum flux expected at the line center, computed from the fluxes in the bracketing sidebands. An index $I > 0$ thus indicates a net absorption. Table 6 lists the line indices, defined by their central bandpass and sidebands. Note that the C IV P-Cygni line index is defined by $I = -2.5 \log(F_{10}/F_{12})$, where $F_{10}$ and $F_{12}$ are the mean fluxes in the bands 10 and 12 respectively.

**Table 5.** Characteristics of the far-UV bands defined by Nandy et al. (1981). The first column gives the identifier of the band (cf. Fig. 13). The initial and final wavelength, and the total width of the bands (in Å) are given in cols. 2, 3, and 4 respectively. Column 5 indicates whether the band is a continuum sideband (C) or the actual bandpass with the line feature (L).

| ID | $\lambda_i$ | $\lambda_f$ | $\Delta\lambda$ | C/L | Comments |
|----|-------------|-------------|------------------|-----|----------|
| 54 | 1685 | 1715 | 30 | C | |
| 56 | 1905 | 1935 | 30 | L | Fe III |
| 57 | 2185 | 2215 | 30 | C | |

Following Nandy et al. (1981), the S1920 line index is determined by $I_i = -2.5 \log(2F_\lambda^i / (F_\lambda^l + F_\lambda^r))$, where $F_\lambda^l$ and $F_\lambda^r$ are the total fluxes in the left and the right sideband respectively (cf. Table 5). The UV line indices derived from our models are plotted in Fig. 14 as a function of the color $(2600 - V)$, which serves as a temperature

**Table 6.** Far-UV spectral line indices

| Feature name | Central wavelength [Å] | Bandpass | Sidebands |
|---|---|---|---|
| Indices from Fanelli et al. (1992): | | | |
| BL 1302 | 1302 | 3 | 2, 4 |
| Si IV 1397 | 1397 | 5 | 4, 8 |
| BL 1425 | 1425 | 6 | 4, 8 |
| Fe V 1453 | 1453 | 7 | 4, 8 |
| C IV abs | 1540 | 10 | 9, 13 |
| C IV cen | 1550 | 11 | 9, 13 |
| C IV emm | 1560 | 12 | 9, 13 |
| C IV P-Cygni | | 10 | 12 |
| BL 1617 | 1617 | 14 | 13, 16 |
| BL 1664 | 1664 | 15 | 13, 16 |
| BL 1719 | 1719 | 17 | 16, 18 |
| BL 1853 | 1853 | 19 | 18, 20 |
| Additional UV line indices: | | | |
| S 1920 | 1920 | 56 | 54, 57 |

indicator and allows a direct comparison with the observations given by FCBW.

### 4.1. Quantitative comparison of CNO and Si lines

First we shall study the four indices which describe the C IV resonance line. The first three indices (centered at 1540, 1550, and 1560 Å) are measuring the strength of the blueshifted absorption, the central absorption, and the redshifted emission respectively. The fourth or 'P-Cygni' index measures the ratio of the emission to the absorption, characterizing the net strength of the wind feature. The most sensitive indices to quantify the P-Cygni line are clearly the blueshifted absorption and the P-Cygni index, which show variations of up to 1 mag over a small range of $(2600 - V)$. As is clear from Fig. 14, our models reproduce both the magnitude and the trends of these C IV indices well. However, as already mentioned above, the ZAMS models predict a too weak C IV $\lambda$ 1550 line. This is clearly an indication that in the hottest models, where C V is predicted as the dominant ionization stage, the carbon ionization seems to be overestimated. For the coolest supergiant models we predict slightly less blueshifted absorption (1540) and too much redshifted emission (1560). This may also be due to an overestimated carbon ionization balance.

For silicon, FCBW introduced the Si IV 1397 index positioned to measure the absorption trough of the Si IV resonance doublet. Its strong luminosity dependence is well known (e.g. Walborn & Panek 1984). Figure 14 shows a systematic shift of the Si IV index. The models have systematically too much flux in the Si IV band with respect to the sidebands. This has to be taken with caution since, dict a too strong O V $\lambda$ 1371 line, which reduces the flux in sideband 4 used to derive the Si IV 1397 index. Artificially eliminating the O V blueshifted absorption leads to an increase of the Si IV 1397 index by $\sim 0\overset{m}{.}15$–$0\overset{m}{.}2$ for the hottest models. In addition, one must consider that for the early type main sequence stars which do not show the Si P-Cygni feature, the 1397 band is also strongly populated by numerous Fe V blends (Bruhweiler et al. 1981). This may also explain part of the shift between our models and the observations. In view of the above, we conclude that the shift in the Si IV 1397 index is most likely not related to an error in the prediction of the silicon ionization structure.

A comparison of the actual profile shape of Si IV with the atlas of Snow et al. (1994) indeed reveals that the observations are well reproduced. Especially the temperature, below which the Si IV line shows a P-Cygni profile is correctly predicted for each track. Moreover, the trend of an increased strength of the absorption trough of the P Cygni line with decreasing temperature is recovered.

The BL 1302 feature is produced by several lines of Si II, Si III, and O I (Fanelli et al. 1992, Shore & Sanduleak 1984), while BL 1425 is due to a blend of Fe V, C III, and Si III. Given the fact that several elements contribute to the corresponding blends, there seems no simple interpretation to explain their behaviour and the small differences with our models. The remaining indices (BL 1664, 1719, and 1853) show no important trends for the hot stars considered in this work. The model predictions for these indices are in tentative agreement with the observations.

### 4.2. Quantitative comparison of Fe features

Of particular interest are line indices which are indicators of the strength of iron lines, since these lines are the most important contributors to blanketing as well as to the radiative force (e.g. Abbott 1982). The models show a strong increase of the metal lines along the evolution from the ZAMS towards the cooler supergiant phase. A very prominent Fe III feature appears at $\lambda \sim 1920$ Å in the late O to early B supergiants. This feature will be investigated in detail in Sect. 4.2. First, we discuss the diagnostics of Fe IV & V.

The Fe V 1453 index shown in Fig. 14 is a measure of a blend of several Fe V lines (cf. Nandy 1977). Its observed strength is $\lesssim 0\overset{m}{.}2$. As shown by the observations of FCBW, and apparent in Fig. 14, this index shows a strong temperature dependence. While our coolest models approximately agree with the observations, we seem to predict too weak a Fe V 1453 absorption for the less evolved models. However, as for the Si IV line discussed above, this index is very sensitive to the predicted O V line strength: By artificially eliminating the O V contribution we find that the indices which use sideband 4 (among which Fe V 1453) may be too low by $\sim 0\overset{m}{.}1$. Correcting

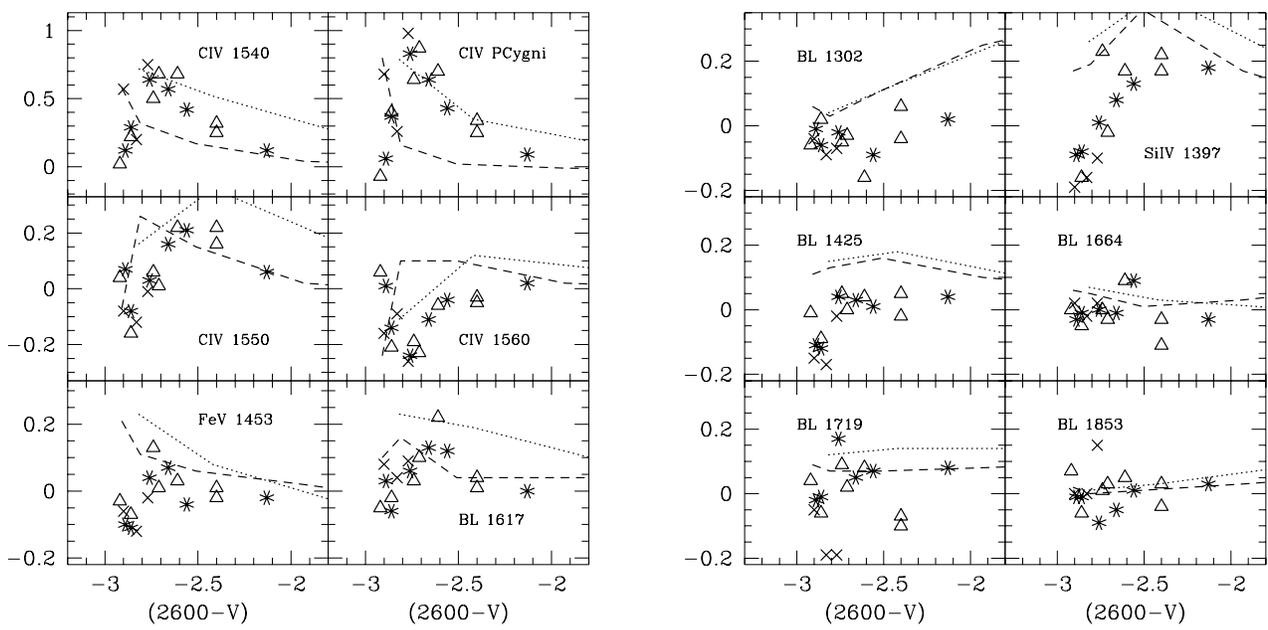

**Fig. 14.** UV line indices in magnitudes as a function of the $(2600 - V)$ color, which serves as a temperature indicator. The symbols (cf. Fig. 1) denote the values predicted by our models. The dashed and the dotted line indicate the mean values derived by FCBW for Dwarfs (LC V) and Supergiants (I) respectively

for this effect a difference of up to $0\overset{m}{.}1$ remains. We therefore conclude that our Fe V 1453 index may point to a somewhat underestimated Fe V ionization.

Another of the FCBW indices measures the contribution of several Fe lines. The BL 1617 absorption is attributed to a blend of Fe IV and Fe V lines (cf. Dean & Bruhweiler 1985). The observations show a clear luminosity separation. Our models predict a behaviour similar to the Fe V 1453 index.

### 4.2.1. The Fe III feature at 1920 Å

As mentioned above, a striking metal line feature around 1920 Å appears in our late O and early B type models. This broad absorption feature, which we shall refer to as the '1920 feature', is due to important blocking of several Fe III lines. It has already been identified in TD1 observation of B type stars (Thomson et al. 1974, Nandy 1977). Subsequently the 1920 feature was shown to increase from late to early B stars (Swings et al. 1976), and to decrease again for O stars (Nandy et al. 1981). To quantify this absorption feature we have computed a photometric measure of its strength, S1920 defined by Eq. 4 and Table 5. It measures the depression in a 30 Å band centered at 1920 Å with respect to continuum points chosen at 1700 and 2200 Å.

In Fig. 15a, we plot the S1920 index derived from our models as a function of effective temperature. In our models the 1920 feature appears first for temperatures $T_{\rm eff} \lesssim$ 33 kK. For these temperatures, we find a strong correlation of its strength with $T_{\rm eff}$. Down to our coolest model with $T_{\rm eff} \sim 18$ kK, the S1920 index increases steadily by up to $0\overset{m}{.}6$–$0\overset{m}{.}8$ with decreasing temperature, as expected by the progressive decrease of the Fe ionization. Note, that in Fig. 15a we have added the value of the S1920 index derived from a model of the Galactic Center He I emission line star AF (Ofpe/WN9; parameters from Najarro et al. 1994, model $B_{\rm IF}$). It shows the strongest 1920 feature of all models calculated so far, and appears as a natural continuation of the early B hypergiants. First, this indicates that if the derived stellar parameters are correct, the effects of line blanketing are expected to be particularly strong. Secondly, this may also be an additional hint showing the intimate relation between LBV's (cf. P Cygni) and Ofpe/WN9 stars. Both points will be discussed in a forthcoming paper.

To allow for a comparison between the predicted Fe III feature and observations we have plotted the S1920 index as a function of absolute visual magnitude in Fig. 15b. The corresponding observations are given by Nandy et al. (1981, their Fig. 8). We note that the strength of the predicted 1920 feature agrees well with the observations, and we correctly reproduce the observed trend of increasing strength from O to early B stars.

To compare the predicted $T_{\rm eff}$ trend with observations, we have derived the S1920 index from IUE spectra for several extensively studied dwarfs and supergiants with relatively low reddening and well determined tem-peratures. All spectra were dereddened with the Seaton (1979) law using the following color excesses. For 9 Sgr: $E(B-V) = 0\overset{m}{.}33$ (Fanelli et al. 1992), $\zeta$ Ori A: $E(B-V) = 0\overset{m}{.}06$, $\kappa$ Ori: $E(B-V) = 0\overset{m}{.}05$ (both Bastiaansen 1992), and P Cyg: $E(B-V) = 0\overset{m}{.}43$ (Lamers et al. 1983). The temperatures are from Bohannan et al. (1990), Voels et al. (1989), Lennon et al. (1991), and Lamers et al. (1983; cf. Pauldrach & Puls 1990). Since it might be argued that the luminous blue variable P-Cygni is a special case, which may not be well represented by our models, we have extended our comparison at low $T_{\rm eff}$ by adding observations of objects that have no individual $T_{\rm eff}$ determination. We use two B2Ia supergiants $\chi^2$ Ori and 10 Per. From the spectral type, Lennon et al. (1993) assign $T_{\rm eff} \sim 17.5$ kK to these stars, while the Schmidt-Kaler (1982) calibration gives $T_{\rm eff} = 18.5$ kK. The $E(B-V)$ color excesses were obtained from Geneva photometry (Rufener 1988) and the intrinsic colors from Cramer (1993). The resulting S1920 indices are $0\overset{m}{.}49$ and $0\overset{m}{.}50$ mag for $\chi^2$ Ori and 10 Per respectively, which agrees well with the value determined for P-Cygni. The observed S1920 indices, plotted in Fig. 15, are found to confirm the predicted strength and the temperature dependence of the Fe III feature.

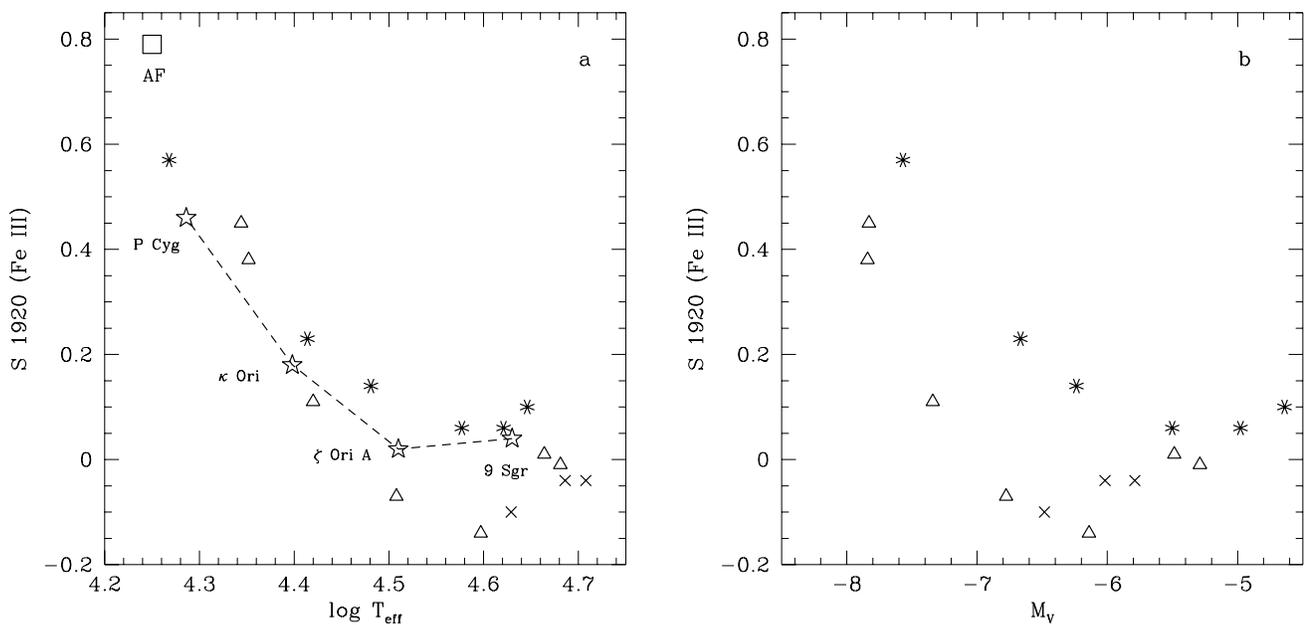

**Fig. 15. a** Predicted line index (in mag) measuring the Fe III 1920 Å feature as a function of $\log T_{\rm eff}$. The same symbols as in Fig. 1 are used. Also shown is one model for the Of/WN Galactic Center star (AF). The dashed line connects observations for stars with well determined $T_{\rm eff}$. Note the strong temperature dependence of S1920, which is confirmed by the observations.
**b** Same as panel a, but shown as a function of $M_V$. This figure can directly be compared with observations of OB supergiants of Nandy et al. (1981)

Apart from showing a temperature trend, one could also expect the Fe III blanketing feature to depend on the wind density. We note however, that although the models with the strongest S1920 index have terminal velocities roughly a factor of two larger than those of $\chi^2$ Ori and 10 Per (Prinja et al. 1990), this effect is compensated by the larger mass loss rate of the models. The comparison of the predicted 1920 feature with the observations is thus well posed.

The above results show that the calculated ionization of Fe agrees well with the observations. Since Fe is the dominant contributor to line blanketing this provides a strong support for our models. In conclusion, the 1920 Å feature provides an excellent test for atmosphere models at $T_{\rm eff} \lesssim 33$ kK. For hotter stars however, where higher ionization stages of Fe are dominant, we must conclude that it is more difficult to obtain constraints of similar quality on the behaviour of Fe.

## 5. Evolution of the optical and IR line spectrum

The most important spectral signatures of OB stars in the optical and IR are the hydrogen and helium lines (see Walborn & Fitzpatrick 1990, Hanson & Conti 1994). To illustrate the predictions made by our models we present in Fig. 16 the evolution of some important H and He lines in the 3500 Å to 4.4 μm interval for the H-burning phase of the 60 $M_\odot$ model. The lines have been obtained from the line blanketed non-LTE calculations assuming a total thermal plus turbulent broadening of 15 km s$^{-1}$. In the present treatment, our predictions should be reliable for

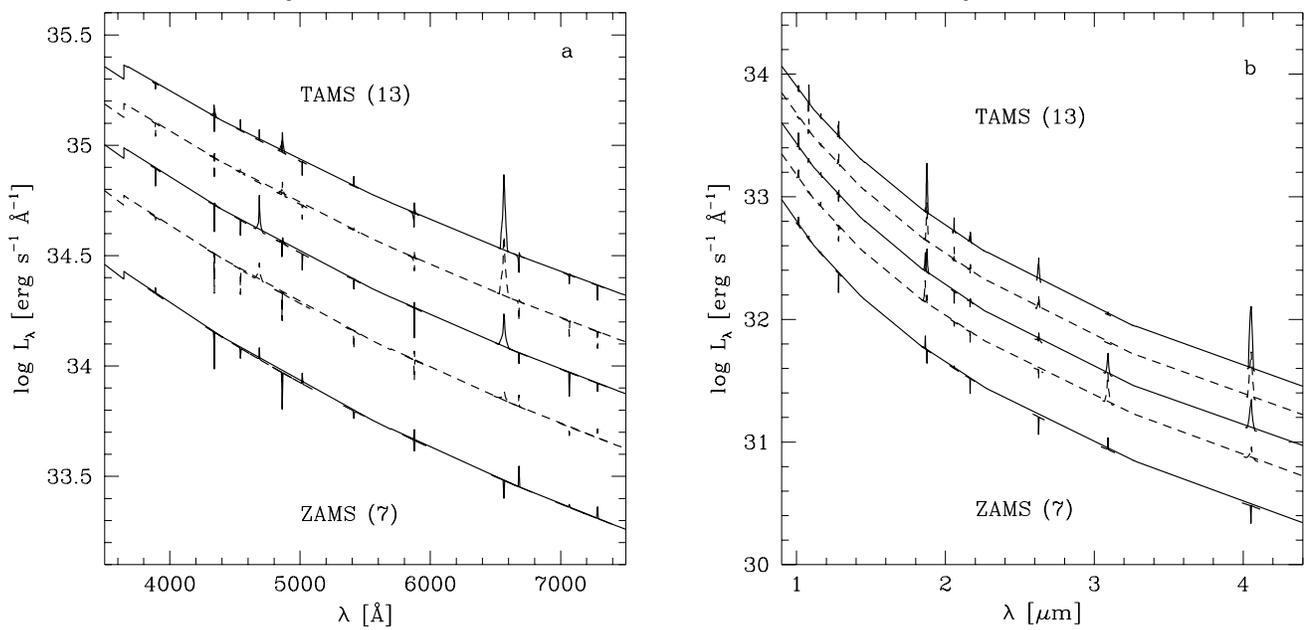

**Fig. 16. a and b.** Predicted line blanketed non–LTE spectral evolution of some important hydrogen and helium lines of a 60 $M_\odot$ star during the MS. Lineprofiles have been superposed to the continuum fluxes. The units and the models shown are the same as in Fig. 2. **a** Visible spectrum. The strongest line plotted is H$\alpha$ ($\lambda$ 6562.8 Å), which changes from absorption to emission with increasing mass loss going from the ZAMS to the TAMS.
**b** IR spectrum covering the IJKL bands. The strongest lines are Pa$\alpha$ $\lambda$ 1.8751 $\mu$m, Br$\beta$ $\lambda$ 2.6252 $\mu$m, and Br$\alpha$ $\lambda$ 4.0512 $\mu$m, which show a similar behaviour as H$\alpha$ depending on the mass loss rate wind lines, while pure absorption lines are less reliable because of the use to the Sobolev approximation and the neglect of pressure broadening (cf. de Koter et al. 1993). Lines contributing to blends are plotted separately. The absence of certain lines is due to the choice of the atomic models.

The most striking change in the optical spectrum is the strong increase of the H$\alpha$ emission towards the end of the MS evolution. This basically reflects the increasing mass loss (see Paper I), which conversely can be used to determine mass loss rates from H$\alpha$ (see below). The strongest lines in the IR region shown here are from the Paschen and Brackett series, which show a similar increase with mass loss as H$\alpha$. This will be discussed in Sect. 5.1. Essentially all predicted Br$\alpha$/Pa$\alpha$ equivalent width ratios (neglecting blends; cf. below) are between 2.8 and 3.8. Only for the models with the weakest emission this ratio is found to increase considerably (up to $\lesssim$ 8).

Of particular interest is also the spectrum in the K band. In this spectral region hot stars in highly obscured star forming regions can be observed. A preliminary spectral classification of OB stars in the K band has been presented by Hanson & Conti (1994). Their observations reveals four main atomic transitions, due to hydrogen, helium, and nitrogen. Here we present the predicted H and He lines in the 2 to 2.25 $\mu$m interval, i.e. Br$\gamma$ (calculated without the He I (7-4) blend), He I 2.058, and the (10-7) He II 2.1885 transition. The observed He I triplet line at 2.113 was not included in our line blanketed non–LTE calculations. The theoretical line profiles obtained for the main sequence evolution of the 40 and 60 $M_\odot$ track are plotted in Fig. 17. Since pressure and rotational broadening are not taken into account in the present calculations the absorption lines are only indicative. We simply plot a triangle profile with a constant width ($\pm 0.1$ in $\Delta v/v_0$) and the relative flux at line center for pure absorption lines. Their interest is to illustrate the relative changes during the main sequence evolution.

Qualitatively, the evolution of the illustrated lines is identical on the two tracks. With the progressive temperature decrease and the mass loss increase, the absorption in the He I and the Br$\gamma$ line gets progressively weaker and switches to emission for the most evolved MS models. The He II line is very weak. It is strongest in ZAMS models, where it is found in absorption.

A comparison with the observations from Hanson & Conti (1994) yields the following: The strongest line in dwarf models (1 & 2 and 7 & 9) is Br$\gamma$, He II 2.189 may show absorption while He I 2.058 is very weak or absent. This qualitatively agrees with the observations of luminosity class V objects. Our supergiant models (5 & 6 and 11 & 13) show for the Br$\gamma$ and the He I line a net emis-sion with an equivalent width ratio of Br$\gamma$/He I $\sim$ 0.5 to 4. The strongest emission is $W_{\rm eq} \sim$ -13. and -4.4 Å for Br$\gamma$ and He I 2.058 respectively. From these emission lines the model predictions close to the TAMS are thus more similar to late OIf types than to early B supergiants (see Hanson & Conti). Although very weak, the He II 2.189 line is predicted in emission contrary to the observations showing He II in absorption in Of supergiant spectra. We have assigned spectral types of $\sim$ B0-2 I to the TAMS stars, using a nearest neighbour search in the $T_{\rm eff}$–$M_{\rm bol}$ plane. However, if one performs a search in $\log T_{\rm eff}$– $\log L$, one obtains spectral type O7I to O9.5I. The comparison with late OIf stars is therefore justified.

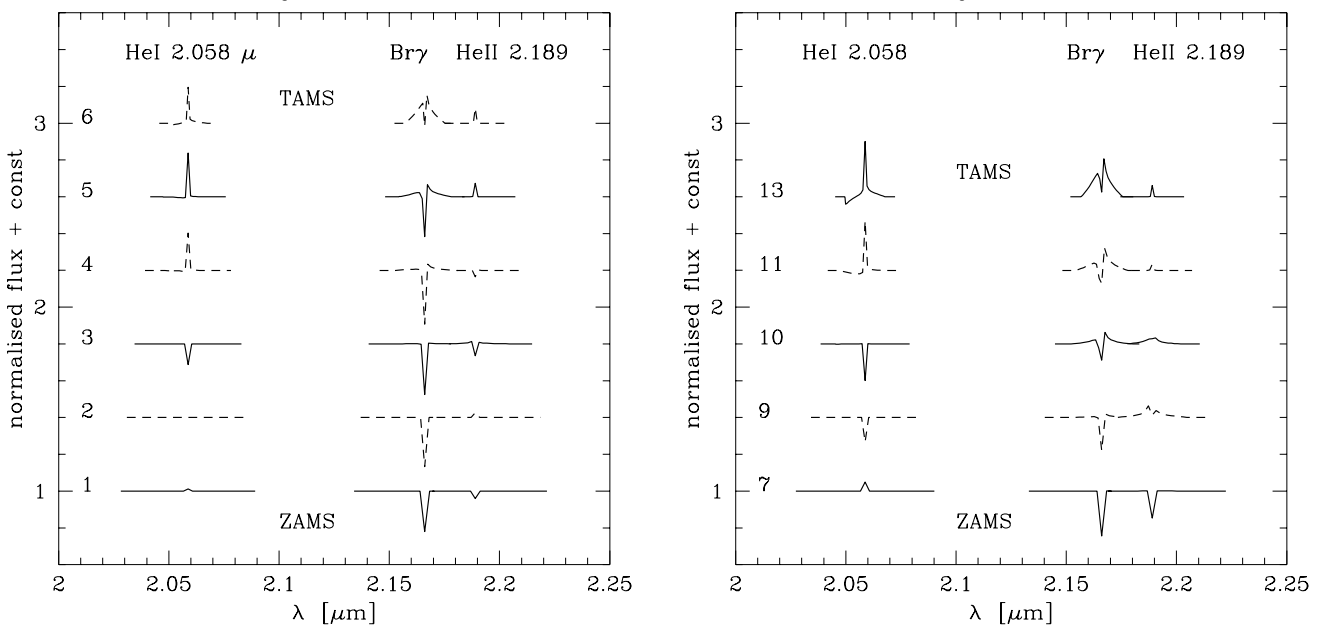

**Fig. 17. a** Predicted K band line profiles for the main sequence evolution of the 40 and **b** 60 $M_\odot$ tracks. Shown are the He I 2.058, Br$\gamma$ and He II 2.1885 lines. The shift between the individual models is 0.4. The model number is according to Table 1. The results for model 8 and 12, which are not shown, are very similar to 7 and 13 respectively

We conclude that the discussed K band lines of H and He are in tentative agreement with the general observed behaviour. This illustrates the capability to reproduce near IR lines with our non–LTE models. More detailed work will be required to allow determinations of stellar parameters of OB stars, based on IR spectra only.

### 5.1. Hydrogen recombination lines as mass loss indicators

#### 5.1.1. Optical spectrum: H$\alpha$

The mass loss rates of hot stars may be determined using the H$\alpha$ recombination line. Given the strong H$\alpha$ emission in hot stars, this technique can readily be applied to extragalactic objects that are not detectable with other techniques such as radio or sub-millimeter measurements. The method has been developed by Leitherer (1988), Scuderi et al. (1992), and Lamers & Leitherer (1993, henceforth LL93).

The procedure can be summarised as follows: Provided the H$\alpha$ luminosity $L({\rm H}\alpha)$, i.e. the distance is known, one subtracts a certain photospheric contribution from the observed H$\alpha$. Assuming an optically thin line and an isothermal wind one can then derive the mass loss rate provided additional stellar parameters (i.e. radius, temperature, wind velocity) are known. An uncertainty in deriving $L({\rm H}\alpha)$ is the assumption of an underlying photospheric absorption profile. The unknown strength of the He II $\lambda$ 6560 blend (cf. LL93), which also contributes to the final observed H$\alpha$ line may also introduce a source of error. More refined procedures, which avoid these assumptions were recently presented by Puls et al. (1995). The important drawback of all these methods is that they require knowledge of photospheric parameters (in any case $T_{\rm eff}$, $\log g$), distance (to derive $R_\star$) and terminal velocity (e.g. from UV observations).

For applications where the above requirements cannot be fulfilled it may be useful to provide a simpler procedure to derive $\dot M$. To this end, we first present a straightforward fit formula to our atmospheric model results, which allows to derive the mass loss rate from a sole H$\alpha$ emission equivalent width measurement. Such a relation is particularly useful for stars with unknown distance and/or unknown wind velocities. In a second step we present an improved relation, taking into account the stellar parameters.

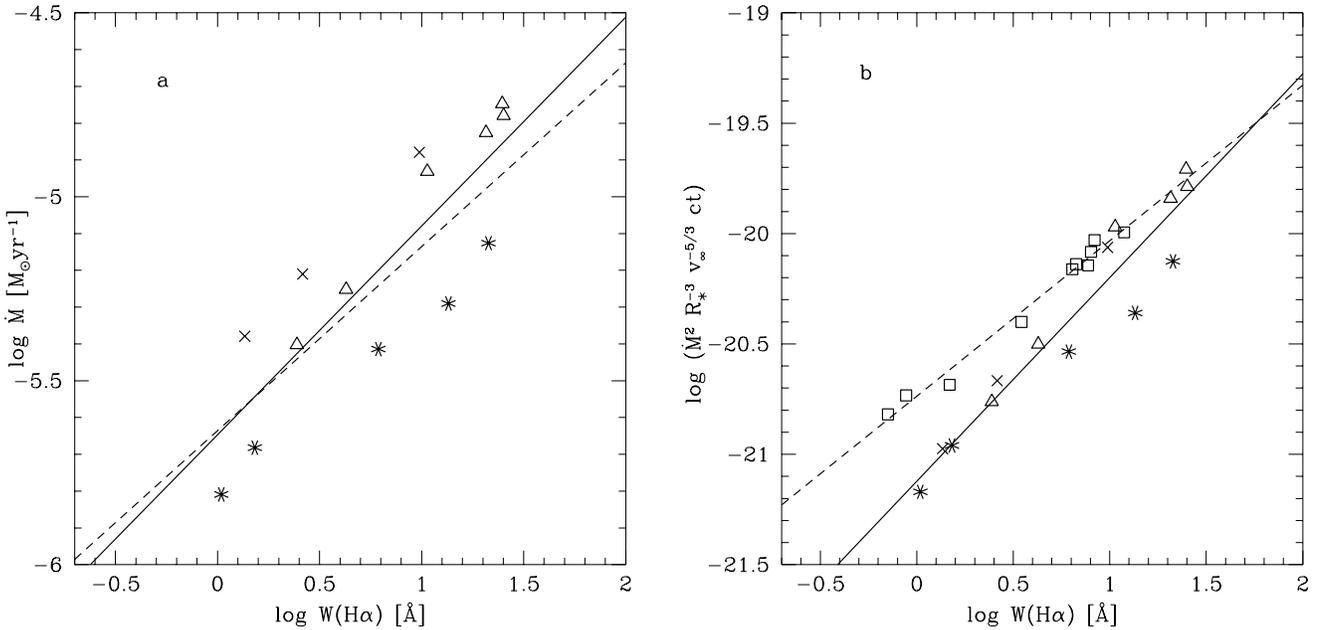

**Fig. 18. a** Mass loss rate as a function of predicted $\log W(H\alpha)$ (emission) including the He II $\lambda$ 6560 blend (same symbols as in Fig. 1; 3 additional models, cf. text). The solid line is a fit to the data (Eq. 1), the dashed line is the relation from Gabler et al. (1990).
**b** Generalised curve of growth for $H\alpha$ obtained from our models (same symbols as in panel a). Shown is $\tilde{Q}^2$ as a function of $\log W(H\alpha)$ (emission). The solid line shows a linear fit to the data, which is used to derive the coefficients $a$ and $b$ used in Eq. 3. Squares show the results from the unified models of Puls et al. (1995, Table 3), which are fitted by the dashed line

In Fig. 18a, we have plotted the mass loss rate as a function of the predicted equivalent width of $H\alpha$. In addition to the models from Table 1 we have included three extra models (found at $0.2 < \log W(H\alpha) < 0.4$ Å) to obtain a more regular distribution of $W(H\alpha)$. Since our models treat the entire atmosphere, including the photosphere and the wind, the predicted $H\alpha$ flux can directly be compared to the observations without having to recur to any arbitrary assumptions about a photospheric contribution and the He II blend. Note that we have only shown the predictions for models having net $H\alpha$ emission, since for weaker lines the present treatment may underestimates the photospheric absorption, and is therefore not sufficiently reliable (cf. de Koter et al. 1993). A least square fit to our data yields:

$$\log \dot{M} = (0.568 \pm 0.100) \log W(H\alpha) - 5.647 \pm 0.093 \quad (1)$$

with a residual rms of $\sim 0.17$ dex. Note that we obtain a quite similar behaviour as given by the relation of Gabler et al. (1990, Eq. 4), which is also plotted in Fig. 18.

As is immediately apparent in Fig. 18a, $W(H\alpha)$ is also dependent on other basic stellar parameters than mass loss. For given $\dot{M}$ the $H\alpha$ emission is stronger in less massive stars. Given their lower luminosity the same $\dot{M}$ is found at lower $T_{\text{eff}}$, which explains the major part of the increased $H\alpha$ emission. Puls et al. (1995) found that the $H\alpha$ equivalent width is expected to scale with $\tilde{Q}^2 \equiv \dot{M}^2 R_\star^{-3} v_\infty^{-5/3} ct$, where $ct$ is a 'temperature correction factor' with respect to a given reference temperature. Following these authors we approximate $ct$ by:

$$ct = \left[\exp\left(\frac{c1}{T_{\text{eff}}}\right) - \exp\left(\frac{c2}{T_{\text{eff}}}\right)\right] T_{\text{eff}}^{-3/2}, \quad (2)$$

where $c1 = 5.184$, $c2 = 2.337$, and $T_{\text{eff}}$ is expressed in $10^4$ K. In Fig. 18b, we plot the behaviour of $\tilde{Q}^2$ as a function of $W(H\alpha)$. This shows that at least half of the scatter present on the left panel can be explained by the scaling relation for optically thin emission taking the stellar parameters $R_\star$, $T_{\text{eff}}$ and $v_\infty$ into account[1]. The best fit of $\log \tilde{Q}^2$ vs. $\log W(H\alpha)$, given as the solid line in Fig. 18b, can be inverted to derive $\dot{M}$. This results in:

$$\begin{aligned}\log \dot{M} &= \frac{1}{2}a\left[\log W + \frac{1}{a}\log\left(R_\star^3 v_\infty^{5/3} ct^{-1}\right)\right] + \frac{1}{2}b \\ &\equiv \frac{1}{2}a \log W'(a, W, R_\star, T_{\text{eff}}, v_\infty) + \frac{1}{2}b\end{aligned} \quad (3)$$

where $W$ is in Å, $R_\star$ in units of $R_\odot$, $v_\infty$ in km s$^{-1}$, and $ct$ is given by Eq. 2. This equation defines the rescaled

---
[1] The apparent scatter in our data stems from models with $T_{\text{eff}} \lesssim 32$ kK, which have considerably larger $b_3$ departure coefficients accounting for their 'excess' emission.

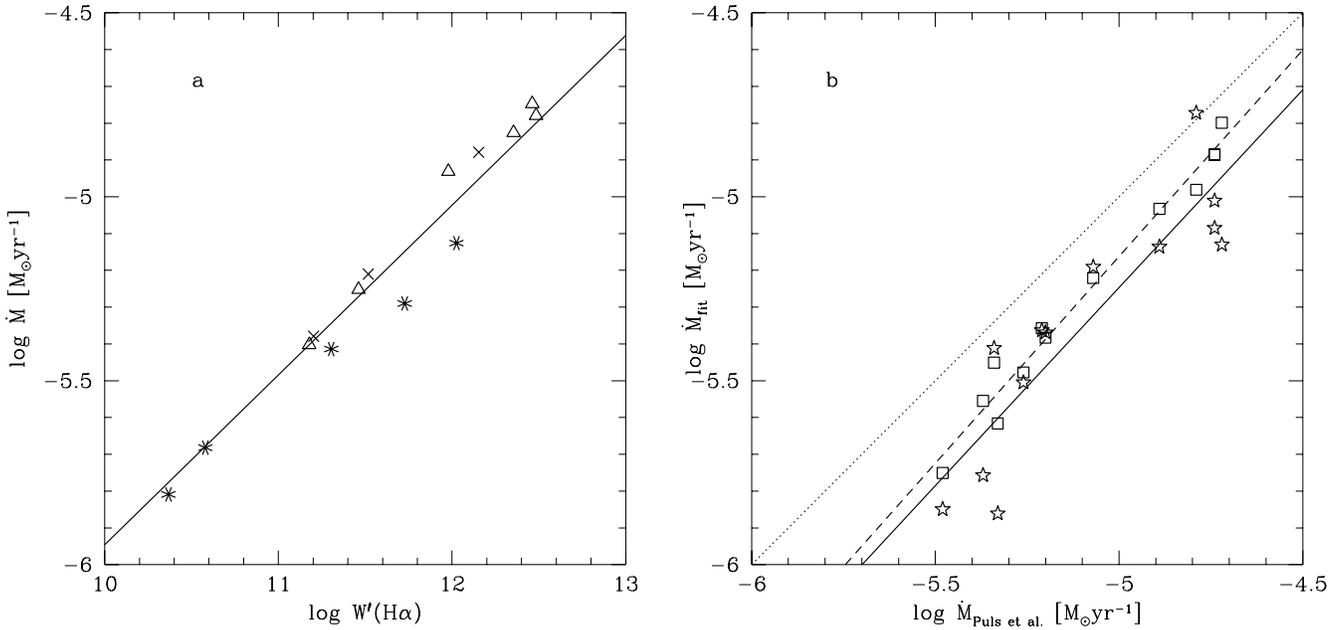

**Fig. 19. a** Mass loss rate as a function of the rescaled Hα equivalent width $W'$ for our models (symbols as in Fig. 18a) The solid line shows the relation given by Eq. 3.
**b** Comparison of mass loss rates ($\dot{M}_{\rm fit}$) using Eq. 1 with $\dot{M}$ determined by Puls et al. (1995, Table 11, col. 3). Plotted are all stars compiled by Puls et al. showing net Hα emission (see LL93). The thin dotted line is the one-to-one relation. Stars show the individual values obtained from Eq. 1 (mean relation: solid line). Squares denote the values obtained from Eq. 3 (mean relation: dashed line) equivalent width $W'$ which also depends on the coefficient $a$. For Hα we obtain $a = 0.923$ and $b = -21.122$. Equation 3 thus allows a more accurate derivation of $\dot{M}$ from the Hα equivalent width provided $R_\star$, $T_{\rm eff}$ and $v_\infty$ are known. The corresponding relation between $W'$ and $\dot{M}$ is shown in Fig. 19a.

In Fig. 18b, we have also plotted the results from the unified models from Puls et al. (squares). A fit to their data (dashed line) yields $a = 0.704$ and $b = -20.737$. Surprisingly, the comparison with our models shows a systematic difference, the models from Puls et al. predicting lower equivalent widths for identical stellar parameters (i.e. same ordinate $\tilde{Q}^2$). Line blanketing, which is the main additional ingredient in our models with respect to Puls et al. only accounts for small changes. This difference must be due to *(a)* differing atmospheric structures in the transition zone between photosphere and wind, which sensitively affects the Hα equivalent width[2] (cf. SS94a), or *(b)* the neglect of pressure broadening, which underestimates the remaining photospheric absorption in our models with relatively weak winds.

To estimate the precision and validity of our relations, we have rederived the mass loss rates for the 13 stars from the sample of LL93 having a net observed Hα emission[3]. The mass loss rate was directly determined from the observed Hα equivalent width using Eq. 1 without any photospheric or other correction. The observational mass loss rates are taken from the new rederivation by Puls et al. (1995, their Table 11, column 3), based on the equivalent widths compiled by LL93. For the individual stars the result is plotted in Fig. 19b, the mean relation being indicated by the solid line. The same is also shown using the stellar parameters from LL93 and Eq. 3 (squares; mean relation: dashed). From this figure we see that our mean relations yield Hα mass loss rates which are on the average 0.1 to 0.3 dex lower than the values derived by Puls et al. This is a direct consequence of the differences illustrated above.

### 5.1.2. The infrared spectrum: Paα & Brα

The IR recombination lines from the Paschen and Brackett series show a similar behaviour as Hα (cf. Fig. 16). With the advent of new IR observations it may be of interest to determine the mass loss rate of OB stars from the strongest emission lines, namely Paα 1.88 μm and Brα 4.05 μm. In Fig. 20 we have used predicted strengths to plot the mass loss rate as a function of equivalent width,

---

[2] Provided the wind velocity structures are identical (∼ same $\beta$) this effect should become more important for weaker winds. This trend is indeed apparent in Fig. 18b.

[3] The smallest equivalent width is of 0.42 Å emission.

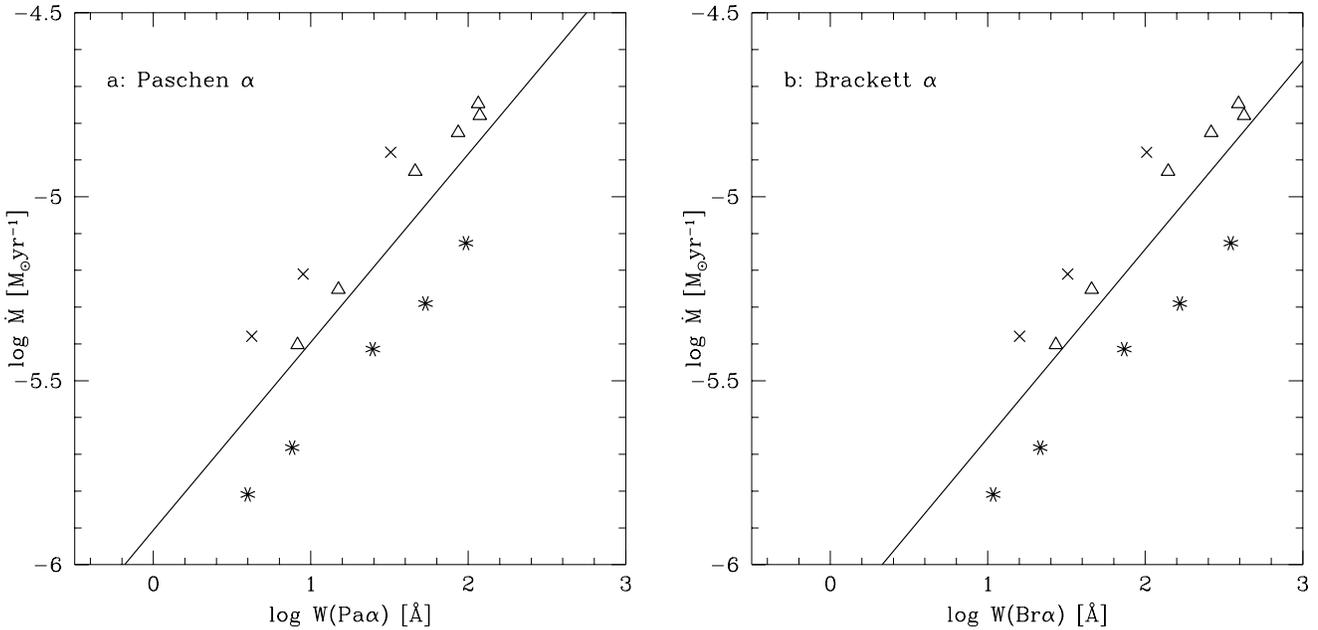

**Fig. 20.** Mass loss rate as a function of the logarithm of the predicted equivalent width (in Å) for Paα (left) and Brα (right). Same symbols as in Fig. 1 with 3 additional models (cf. text). The solid lines show the relations given by Eqs. 4 and 5 respectively

for the same models as for Hα above (again restricted to net emission in the lines). Note that we have not included the He I (4-3) blend in Paα. Depending primarily on $T_{\rm eff}$ and the He abundance, this would lead to a small increase of the total Paα equivalent width and thereby to a flatter relation. Similarly the He II (10-8) transition which blends Brα was not considered.

Figure 20 clearly exhibits the strong dependence of Paα and Brα on $\dot{M}$. The following mean relations are obtained from our model results:

$$\log \dot{M} = (0.511 \pm 0.102) \log W({\rm Pa}\alpha) - 5.907 \pm 0.152 \quad (4)$$

$$\log \dot{M} = (0.513 \pm 0.095) \log W({\rm Br}\alpha) - 6.169 \pm 0.188 \quad (5)$$

The residual rms is 0.18 and 0.17 dex respectively, which is of the same order as for Hα.

In analogy to the case of Hα discussed above, the accuracy of the $\dot{M}$ determination can be improved if, in addition, $R_\star$, $T_{\rm eff}$ and $v_\infty$ are known. For the temperature correction $ct$ (Eq. 2) the constants $c1$ and $c2$ have the following values: $c1 = 2.311$ (1.303) and $c2 = 1.315$ (0.842) for Paα and Brα respectively. The resulting fit values for Eq. 3 are $a = 0.703$ (0.638) and $b = -22.111$ (-22.785). We note that, given the large Paα and Brα emission, these lines should be less affected by the uncertainty found above for Hα from the comparison with Puls et al. the models.

As a first comparison between determinations based from the optical and the IR, we calculate $\dot{M}$ for ζ Puppis (HD 66811). The observed Hα equivalent width of -3.09 Å is taken from the compilation of LL93. From the line profiles of Käufl (1993) we measured Brα equivalent widths of 74 to 101 Å. The derived mass loss rate (in $M_\odot\,{\rm yr}^{-1}$) is $\log \dot{M}$= -5.37 from Hα, and $\log \dot{M}$= -5.21 to -5.14 from Brα, which shows a reasonable agreement between both determinations[4]. From their detailed analysis including a line profile fit of Hα, Puls et al. (1995) derive $\log \dot{M}$ = -5.23, which confirms our value obtained from pure equivalent widths measurements. If, in addition, we take the stellar parameters of LL93 into account and use Eq. 3 with the corresponding constants, we obtain $\log \dot{M}$ = -5.39 from Hα, and $\log \dot{M}$ = -5.31 to -5.27 from Brα again in good agreement with the observed value. More detailed comparisons are beyond the scope of the present study.

## 6. Summary and conclusions

We have presented the spectral evolution of main sequence stars with initial masses $M_i$ = 40, 60 & 85 $M_\odot$, using "combined stellar structure and atmosphere models" (*CoStar*). The first *CoStar* studies of the advanced Wolf-Rayet phases are presented by Schaerer (1995ab). *CoStar* models consistently combine stellar structure calculations with non-LTE line-blanketed model atmospheres. The technical details of *CoStar* and the evolution tracks used in this work are discussed in Paper I of this series. The main goal of this second paper is to confront our models

---

[4] Despite the known He enrichment of ζ Puppis the use of our fit formulæ is justified for this case, since the total Hα equivalent width including the He II 6520 blend is expected to vary by less than ∼ 8%.

a more 'direct' comparison of stellar evolution calculations with observations.

The presented predictions for the spectral evolution cover: *(a)* line blanketed flux distributions from the EUV to the far IR, *(b)* the integrated ionizing fluxes for the hydrogen and helium continua, *(c)* photometry for eleven broad-band filters in the optical and infrared and for four filters in the UV, *(d)* synthetic UV spectra ($\Delta\lambda/\lambda$ up to $10^3$) for the IUE wavelength range from 1000 to 2200 Å, including the dominant wind and metal line features, and finally *(e)* the optical and infrared hydrogen and helium line spectrum.

We have performed these calculations at several points along the main sequence for three different initial masses. They therefore present a homogeneous set of results for a large domain of stellar parameters of massive stars, which will be of value for studies of nebulae encompassing early-type stars and for studies of young stellar populations in general (e.g. in H II regions or starbursts).

At this point we will not summarize all our findings, but point out three interesting results. The first concerns the ionizing spectrum, the second the iron spectrum, and the third is a method to derive mass loss rates from optical and infrared lines.

Comparisons of our *CoStar* ionizing spectra with other calculations based on LTE or non–LTE plane parallel models clearly show the importance of wind effects on the determination of the ionizing flux of O and, at least, early B stars (cf. Gabler et al. 1992, Najarro et al. 1995). This may be of considerable importance for the interpretation of spectra of H II regions and related objects.

Since iron is the main contributor to line blanketing and to the radiative force in the investigated stars, we have paid special interest to diagnostics related to these two fundamental problems. The broad band UV colors are found to be quite sensitive to line blanketing effects. A comparison of these colors with spectrophotometric IUE observations shows an encouraging agreement. A number of metal line features in the spectral range mostly covered by the IUE short wavelength camera are used to investigate the ionization structure of iron. The strongest feature in the spectrum is due to a large number of Fe III lines at $\lambda \sim 1920$Å. The predictions of this feature, which shows a strong temperature dependence, are in excellent agreement with observations. They may even be used as a temperature indicator for late O and early B giants and supergiants ($18 \lesssim T_{\rm eff} \lesssim 33$ kK). Other iron features, such as the relatively weak blend of Fe V lines around 1453 Å, do not provide such strong diagnostic tools. The Fe V 1453 feature even points to a somewhat underestimated iron ionization in our models. In general however, we conclude that the predictions of various investigated spectral features are in reasonable to satisfactory agreement with observations.

to derive simple fit formulae, which allow the determination of the mass loss rate from the equivalent width for emission of each of these three lines. A test of this method using the O4I(n)f star $\zeta$ Puppis indicates that this method yields reliable results.

*Acknowledgements.* We thank Dr. Allan Willis for stimulating discussions and pointing out early UV observations. Dr. Artemio Herrero and David Bersier kindly provided us with results from atmosphere models. We acknowledge Dr. Carmelle Robert for sending us a routine for spectral type determinations. Particular thanks from D.S go to Prof. André Maeder for his continuous encouragement and support. Part of this work was supported by the Swiss National Foundation of Scientific Research and by NASA through a grant to the GHRS science team. This research has made use of the Simbad database, operated at CDS, Strasbourg, France.